\documentclass[
 reprint,
 superscriptaddress,
 nofootinbib,
 amsmath,amssymb,
 aps,prx
]{revtex4-2}

\usepackage{graphicx}
\usepackage{color}
\usepackage{natbib}
\usepackage{epsfig}
\usepackage{float}
\usepackage{xcolor}
\usepackage{soul}
\usepackage{comment}
\usepackage{amsmath,amssymb,amsfonts}
\usepackage{subfigure}
\usepackage{hyperref}
\usepackage[normalem]{ulem}
\usepackage{longtable}

\DeclareUnicodeCharacter{2212}{\textendash}

\newcommand{\apjl}{Astrophys. J. Lett.}%
\newcommand{\aap}{Astron. Astrophys.}%
\begin{document}

\title{Comprehensive survey of hybrid equations of state in neutron star mergers and constraints on the hadron-quark phase transition}

\author{Sebastian Blacker}
\affiliation{Institut f\"ur Kernphysik, Technische Universit\"at Darmstadt, 64289 Darmstadt, Germany}
\affiliation{GSI Helmholtzzentrum f\"ur Schwerionenforschung, Planckstra{\ss}e 1, 64291 Darmstadt, Germany}

\author{Andreas Bauswein}
\affiliation{GSI Helmholtzzentrum f\"ur Schwerionenforschung, Planckstra{\ss}e 1, 64291 Darmstadt, Germany}
\affiliation{Helmholtz Research Academy Hesse for FAIR (HFHF), Campus Darmstadt, 64291 Darmstadt, Germany}

\date{\today}

\begin{abstract}
We perform an extensive study of equation of state (EoS) models featuring a phase transition from hadronic to quark matter in neutron star merger simulations. We employ three different hadronic EoSs, a constant speed of sound parameterization for the quark phase and a Maxwell construction to generate a large sample of hybrid EoS models. We systematically vary the onset density and density jump of the phase transition as well as the quark matter stiffness and simulate binary neutron star mergers to infer how the properties of the phase transition affect the gravitational-wave signal. In total we simulate mergers with 245 different hybrid EoS models. In particular, we explore in which scenarios a phase transition would be detectable by a characteristically increased postmerger gravitational-wave frequency compared to an estimate from the inspiral signal assuming a purely hadronic EoS. We find that the density jump at the transition (latent heat) has the largest impact on the gravitational-wave frequencies, while the influence of the stiffness of quark matter is smaller. We quantify which range of phase transition properties would be compatible with a certain magnitude or absence of the gravitational-wave postmerger frequency shift. By means of these dependencies, a future detection will thus directly yield constraints on the allowed features of the hadron-quark phase transition.
\end{abstract}


\maketitle   

\section{Introduction}
At high densities or temperatures matter is expected to undergo a transition from hadronic to quark matter. Around zero baryon chemical potential, this transition is relatively well explored from collider experiments and lattice quantum chromodynamics (QCD) calculations~\cite{Bazavov:2012vg,Borsanyi:2013bia,HotQCD:2014kol,Andronic:2016nof,HotQCD:2018pds,Pandav:2022xxx,MUSES:2023hyz}. At finite baryon chemical potential, the QCD phase diagram is much less understood as the sign problem currently prevents theoretical calculations in this regime~\cite{Nagata:2021ugx}. It has long been hypothesized that the transition from hadronic to deconfined quark matter could occur at densities reached in the interiors of neutron stars (NSs) and binary neutron star (BNS) merger remnants~\cite{Ivanenko:1965dg,Ivanenko:1969gs,Itoh:1970uw,Chapline:1976gq,Baym:1976yu,Keister:1976gc,Glendenning:1992vb}, see e.g.~\cite{Baym:2017whm,Blaschke:2018mqw,Oertel:2016bki,Alford:2019oge,Orsaria:2019ftf,Hanauske:2019czl,Raduta:2022elz,Bauswein:2022vtq} for reviews. Studying these compact objects and confirming or ruling out the presence of deconfined quark matter therefore provides valuable insights into the QCD phase diagram. 

The transition to deconfined quark matter has been studied in the context of BNS merger simulations and core collapse supernovae employing so-called hybrid equation of state (EoS) models, i.e. models with a hadronic and a deconfined quark phase connected by a phase transition~\cite{Gentile:1993ma,Oechslin:2004yj,Sagert:2008ka,Olson:2016cxd,Fischer:2017lag,Bauswein:2018bma,Most:2018eaw,Hanauske:2019czl,Weih:2019xvw,DePietri:2019khb,Ecker:2019xrw,Blacker:2020nlq,Liebling:2020dhf,Bauswein:2020aag,Bauswein:2020ggy,Bauswein:2020xlt,Prakash:2021wpz,Zha:2021fbi,Tootle:2022pvd,Fujimoto:2022xhv,Jakobus:2022ucs,Fujimoto:2022xhv,Huang:2022mqp,Ujevic:2022nkr,Kedia:2022nns,Haque:2022dsc,Espino:2023llj,Prakash:2023afe,Harada:2023eyg,Largani:2023oyk,Guo:2023som,Kumar:2023qyu,Ecker:2024kzs,Hammond2025,Fujimoto2025}. These studies are limited to a small set of particular EoS models. Typically, the appearance of deconfined quark matter leads to more compact stars or merger remnants and correspondingly to higher postmerger gravitational-wave~(GW) frequencies compared to purely hadronic models. This has been pointed out as a smoking-gun signal to identify the occurrence of the hadron-quark phase transition~\cite{Bauswein:2018bma,Weih:2019xvw,Blacker:2020nlq,Bauswein:2020ggy}. Additionally, remnants with hybrid EoS tend to collapse to black holes earlier~\cite{Most:2018eaw,Bauswein:2020aag,Bauswein:2020xlt,Ecker:2024kzs}.

However, the magnitude of the postmerger GW frequency shift due to the occurrence of deconfined quark matter can vary significantly depending on the properties of the hybrid EoS model. Reference~\cite{Bauswein:2018bma} already indicates a dependence on the latent heat, but because of the limited number of available fully temperature-dependent hybrid EoS models, the systematic dependencies are elusive to date. It is for instance also possible that the properties of the hadron-quark phase transition might be such that they do not lead to a pronounced impact on the postmerger GW frequency and thus the system may be indistinguishable from a purely hadronic one. This ambiguity is commonly refereed to as the masquerade problem~\cite{Alford:2004pf}. This motivates to determine the quantitative dependencies of the postmerger GW frequency shift on the properties of the phase transition and in general the hybrid EoS to understand in which regions of the parameter space the aforementioned degeneracies occur and for which configurations a pronounced signature of the phase transition can be expected. In turn, the properties of the phase transition can be constrained from the magnitude of the frequency shift. 

In this work, we provide for the first time such a quantitative analysis of how different properties of the hybrid EoS impact the postmerger GW signal. For this, we perform merger simulations with three different underlying hadronic EoSs, a constant speed of sound parametrization~\cite{Zdunik:2012dj,Chamel:2012ea,Alford:2013aca} for the quark phase and a Maxwell construction for the phase transition~\cite{Glendenning:1992vb,Hempel:2009vp,Hempel:2013tfa,Constantinou:2023ged}. Additionally, we employ the scheme of Ref.~\cite{Blacker:2023afl} to include thermal effects. By independently varying the onset density, the density jump at the transition and the stiffness of the quark phase we explore how these quantities affect the GW signal and the detectability of the phase transition. For each underlying hadronic EoS, we provide a fit formula that quantifies the impact of the hybrid EoS properties on the dominant postmerger GW frequency. Once postmerger GWs are detectable, this can immediately place constraints on properties of the transition in the high-density and low-temperature region of the QCD phase diagram.

This paper is organized as follows: In Sec.~\ref{sec:methods} we briefly discuss the employed hybrid EoS model, the range of employed parameters and the numerical simulation code. We present the simulation results in Sec.~\ref{sec:results} were we show how different hybrid EoS properties affect the GW signal. We summarize and conclude in Sec.~\ref{sec:summary}.

\section{Methods and setup}\label{sec:methods}
\subsection{Hybrid equations of state}\label{subsec:eos}
To construct different hybrid EoSs, we employ a constant speed of sound parametrization of the quark phase following Ref.~\cite{Chamel:2012ea,Zdunik:2012dj,Alford:2013aca}. This approach is able to reproduce existing microphyiscal models such as NJL based calculations as in~\cite{Shahrbaf:2021cjz} or the DD2F-SF models~\cite{Bastian:2020unt} (see Appendix~\ref{app:eos}) very well. We present the resulting equations for pressure, energy and number density in Appendix~\ref{app:eos}. {We emphasize that our approach is independent of the microphysics of the phase transition and { of} the quark matter EoS, and we do not explicitly distinguish between the chiral phase transition and the onset of deconfinement. For the purpose of this study, the model independence is a particular advantage because it allows to tune the phase transition properties and to build a large set of models, which was not possible by employing full microphysical EoS tables that are not available in large numbers and do not offer the flexibility to adapt the parameters of the phase transition. We note that thermal effects of the EoS are included by an effective scheme 
with the most prominent impact emerging from the temperature dependence of the phase boundaries, which should be considered as additional degree of freedom of the EoS (see~Sect.~\ref{subsec:code}, App.~\ref{app:effpt} and~\cite{Blacker:2023afl}).}
%

At zero temperature, we match {the constant-speed-of-sound} quark matter model to one of three hadronic EoSs, namely DD2F~\cite{Typel:2009sy,Alvarez-Castillo:2016oln}, DD2~\cite{Typel:2009sy,Hempel:2009mc} and SFHo~\cite{Hempel:2009mc,Steiner:2012rk}, using a Maxwell construction~\cite{Glendenning:1992vb,Glendenning:2001pe,Hempel:2009vp,Hempel:2013tfa,Constantinou:2023ged}.
This scheme leads to a density jump at the phase transition. We refer to the onset baryon number density of the transition as $n_\mathrm{on}$ and the size of the density jump as $\Delta n$. The onset baryon number density of the pure, deconfined quark phase $n_\mathrm{fin}$ is then given by $n_\mathrm{fin}=n_\mathrm{on}+\Delta n$.

A specific choice of $n_\mathrm{on}$, $\Delta n$ and a speed of sound $c_s$ in the quark phase together with a given hadronic model completely determines the hybrid EoS at $T=0$. By systematically varying these parameters, we generate an ensemble of different hybrid models.

Observations of massive pulsars indicate that the maximum NS mass is at least $2~M_\odot$~\cite{Antoniadis:2013pzd,Fonseca:2021wxt,Romani:2022jhd}. This poses constraints on combinations of $n_\mathrm{on}$, $\Delta n$ and $c_s$ in the hybrid EoS approach. Additionally, the speed of sound is limited by causality, i.e. it must be below the speed of light.

\begin{figure*}
\centering
\subfigure[SFHo]{\includegraphics[width=0.33\linewidth]{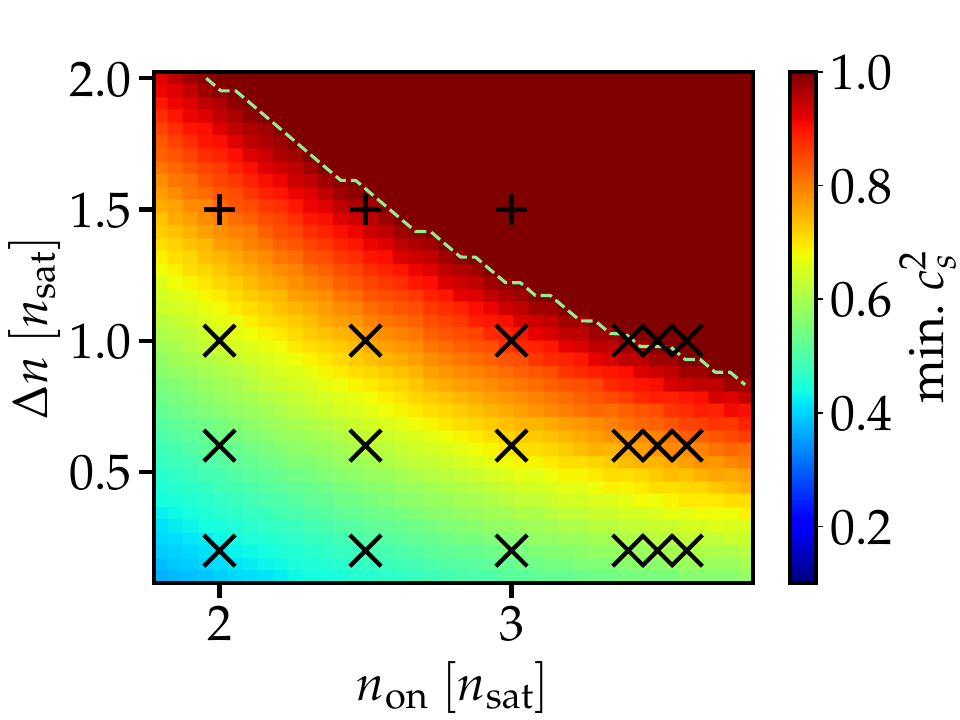}\label{fig:c2_constraints_a}}
\hfill
\subfigure[DD2F]{\includegraphics[width=0.32\linewidth]{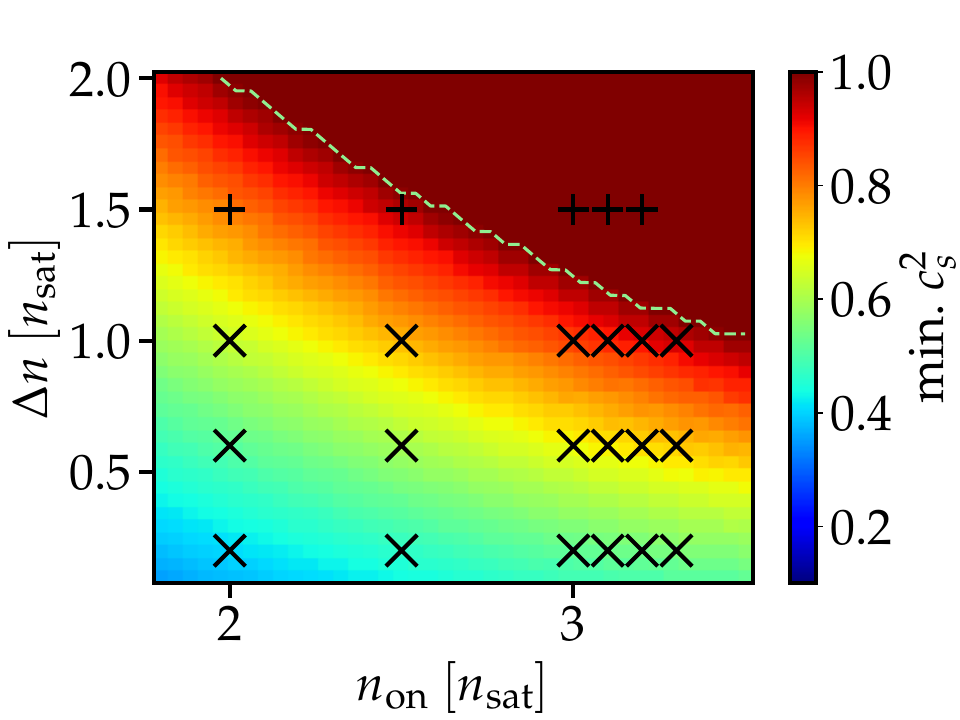}\label{fig:c2_constraints_b}}
\subfigure[DD2]{\includegraphics[width=0.32\linewidth]{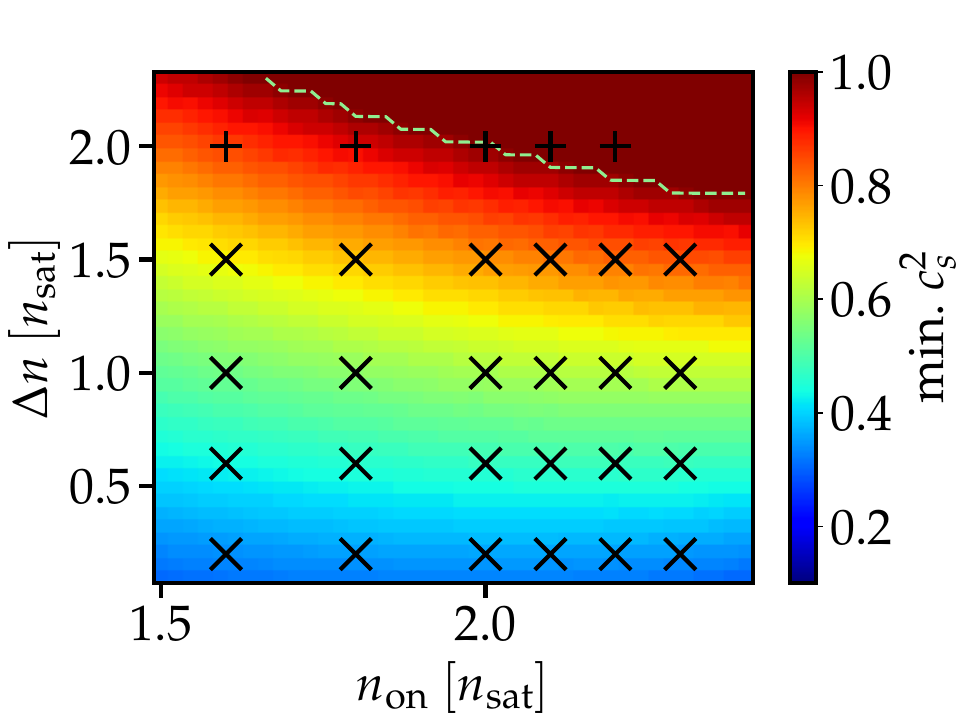}\label{fig:c2_constraints_c}}
\hfill
\caption{Minimum speed of sound squared in the quark phase required to obtain a maximum neutron star mass $M_\mathrm{max}\geq 2~M_\odot$ color-coded as a function of onset density and density jump of the phase transition for hybrid models based on the hadronic SFHo (left panel), DD2F (middle panel) and DD2 (right panel) models. The stiffness of the hadronic models increases from left to right. In each panel the dashed line encloses the region where the requirement $M_\mathrm{max}\geq 2~M_\odot$ cannot be fulfilled. Black symbols mark {EoS models for which we perform simulations of binary NS mergers} in this work with plus signs indicating models that promptly collapse to a black hole at merger for all $c_s^2\leq 1$. For EoSs with the highest simulated $n_\mathrm{on}$ and beyond, the phase transition does not have an impact on the GW signal anymore.}
\label{fig:c2_constraints}
\end{figure*}

We demonstrate the effect these constraints have on the parameter space in Fig.~\ref{fig:c2_constraints} for all three hadronic EoSs (compare e.g.~\cite{Gorda:2022lsk,Christian:2023hez} for recent works on constraints for deconfined quark matter). Since EoSs with larger $c_s$ can support larger maximum masses, we plot the minimum $c_s^2$ necessary to reach a maximum NS mass of $2~M_\odot$ for different combinations of $n_\mathrm{on}$ and $\Delta n$. $n_\mathrm{on}$ and $\Delta n$ are given in units of the nuclear saturation density $n_\mathrm{sat}=0.16~\mathrm{fm}^{-3}$. It is apparent that larger $n_\mathrm{on}$ and $\Delta n$ require a stiffer quark phase to fulfill the pulsar mass constraint. We also see that the DD2 EoS generally permits softer quark matter since it is stiffer than the other two hadronic models. The dark red area enclosed by the dashed line in the top right corner of each panel represents a region of the parameter space, where even the maximum sound speed of $c_s=1$ does not yield $M_\mathrm{max}\geq 2~M_\odot$. Hence, this region is excluded by observations.

The black symbols mark configurations of $n_\mathrm{on}$ and $\Delta n$, {i.e. EoS models, for which we perform NS merger simulations} in this work. For each pair, we pick different values of $c_s^2$ ranging from $1$ down to the lowest values indicated in Fig.~\ref{fig:c2_constraints} for the given $\Delta n$ and $n_\mathrm{on}$ and simulate mergers of two $1.35~M_\odot$ stars. Crosses mark configurations allowing for (temporary) stable merger remnants and the occurrence of a strong postmerger GW signal while plus signs depict models where we find prompt black-hole formation at merger even for maximum quark matter stiffness. We also include some configurations with maximum NS masses slightly below $2~M_\odot$ to cover a broader region of the hybrid model space. We do not consider parametrizations with larger $n_\mathrm{on}$ as we find that for those the phase transition no longer impacts the gravitational wave signal since the produced amount of quark matter becomes too small to noticeably affect the structure of the remnant.

Note that all generated hybrid EoSs are barotropic models, where the pressure solely depends on the baryon density and the composition is always assumed to be in neutrinoless $\beta$-equilibrium. 

\begin{figure*}
\centering
\subfigure[]{\includegraphics[width=0.33\linewidth]{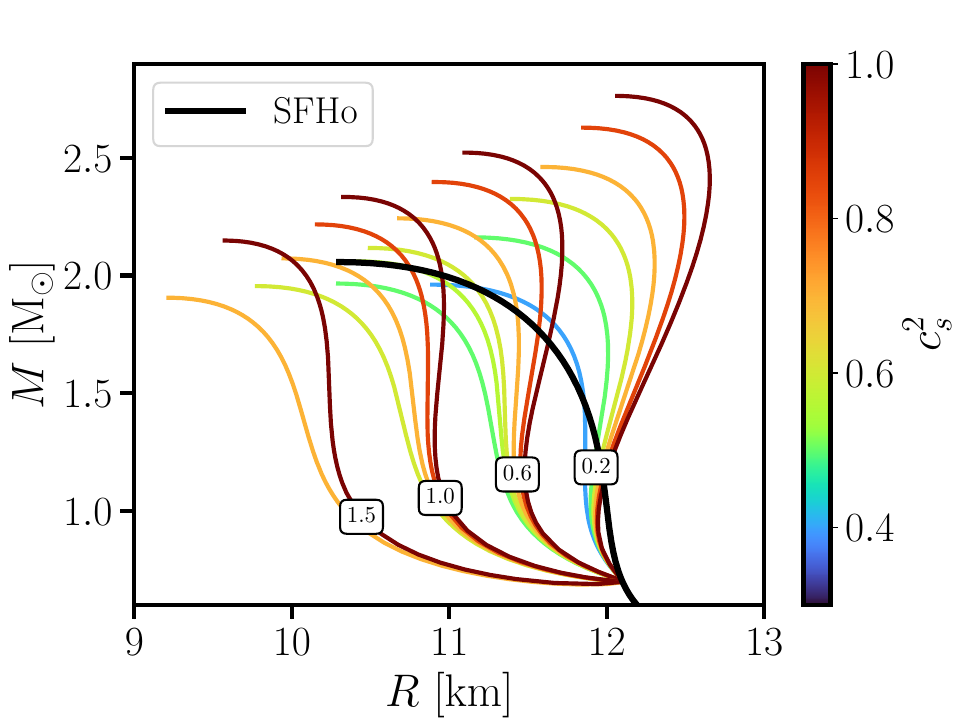}\label{fig:tov_a}}
\hfill
\subfigure[]{\includegraphics[width=0.32\linewidth]{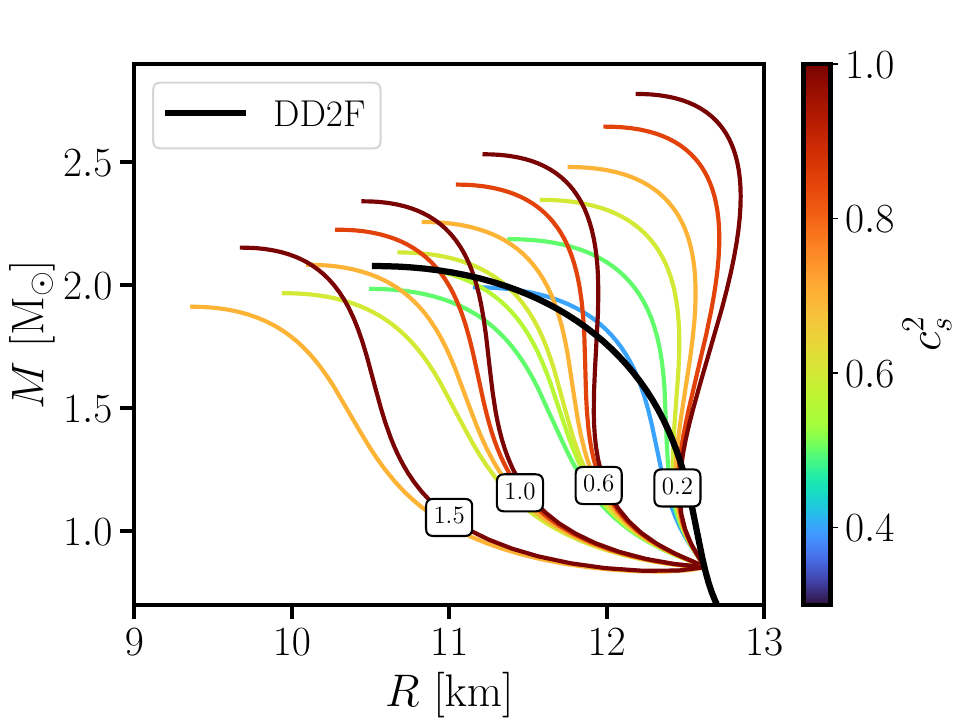}\label{fig:tov_b}}
\subfigure[]{\includegraphics[width=0.32\linewidth]{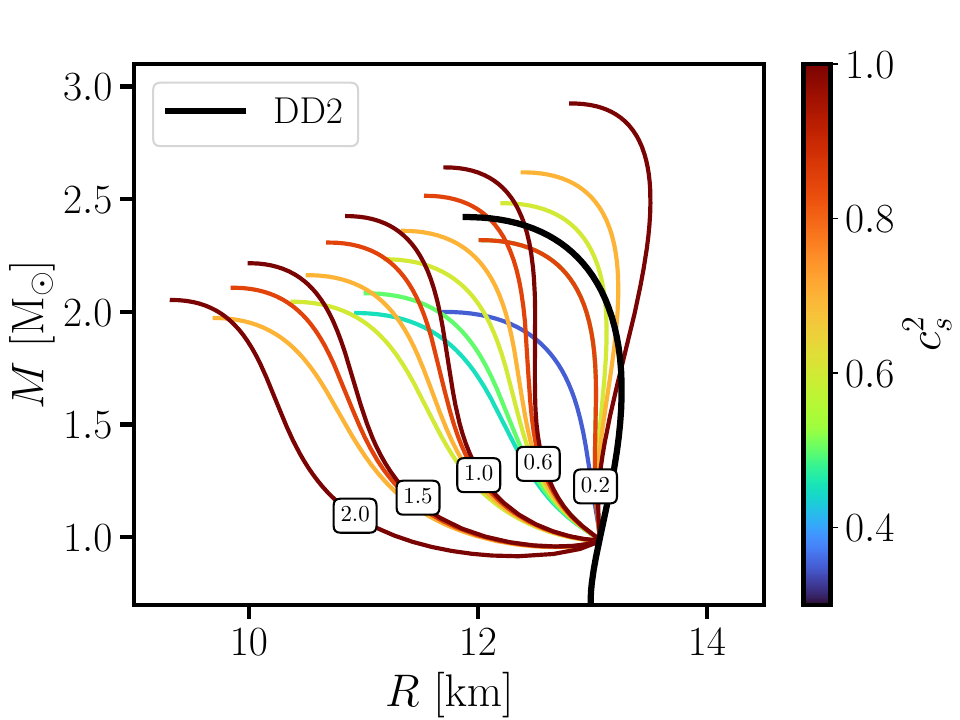}\label{fig:tov_c}}
\\
\subfigure[]{\includegraphics[width=0.33\linewidth]{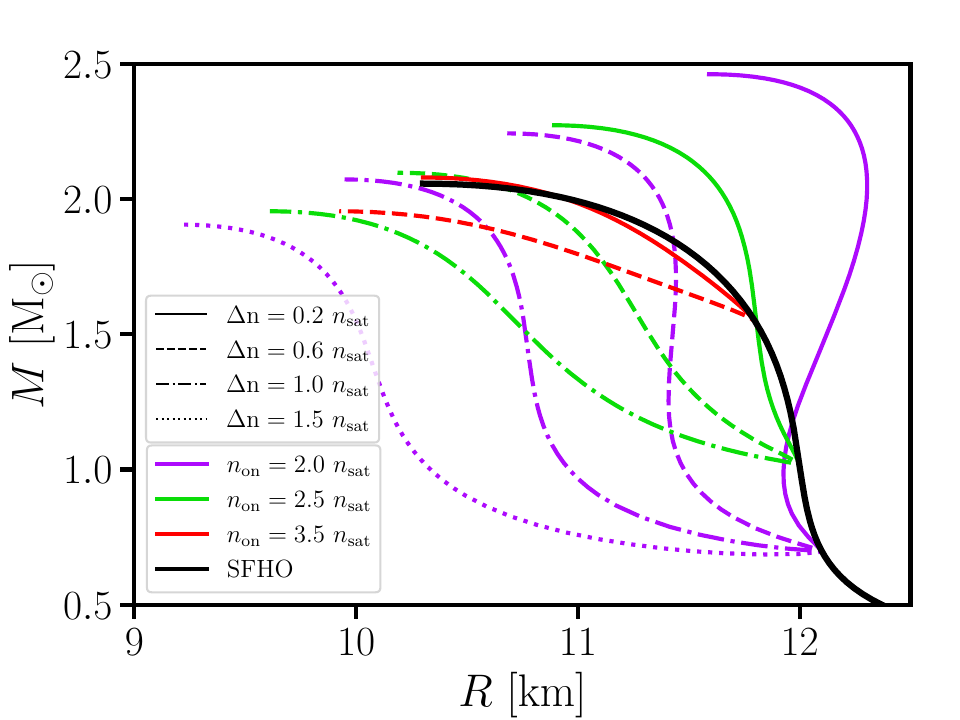}\label{fig:tov_d}}
\hfill
\subfigure[]{\includegraphics[width=0.32\linewidth]{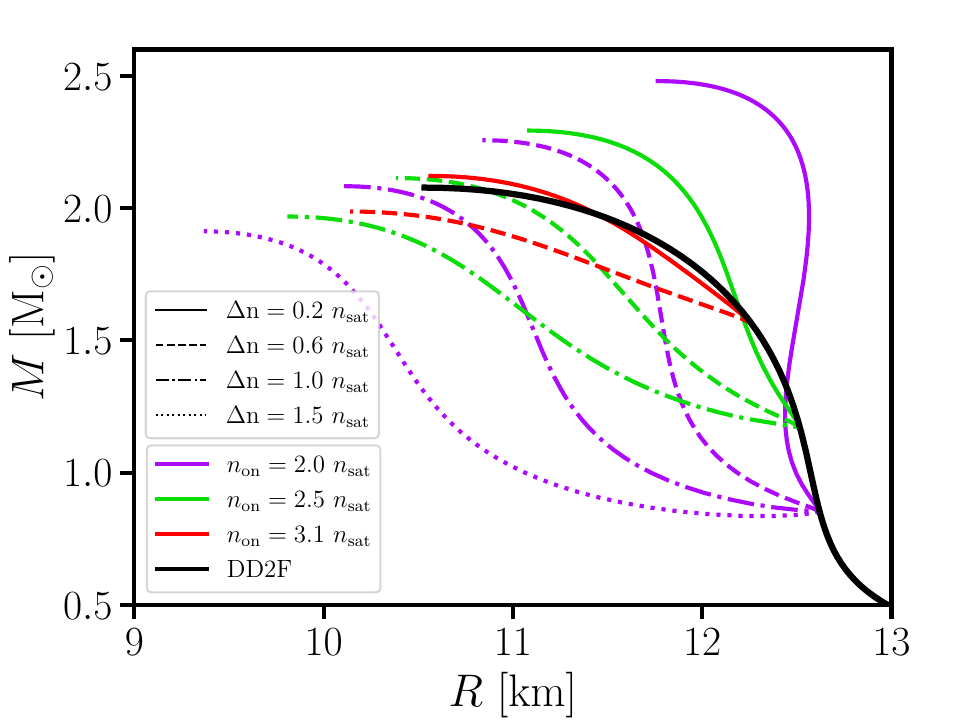}\label{fig:tov_e}}
\subfigure[]{\includegraphics[width=0.32\linewidth]{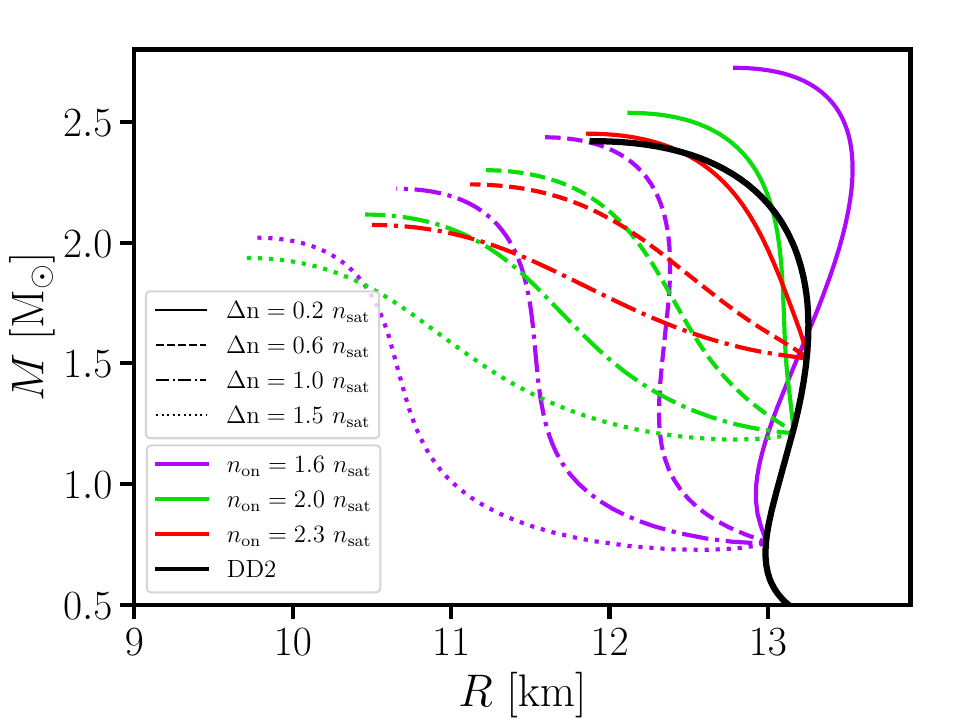}\label{fig:tov_f}}
\hfill
\caption{Mass-radius curves of cold neutron stars for a selection of hybrid EoSs we employ in this study based on a constant speed of sound model (colored). Each column shows results from hybrid models based on a different underlying hadronic model (black). The top row displays curves of hybrid models with a single onset density of the phase transition {of $2\times n_\mathrm{sat}$ (SFHO), $2\times n_\mathrm{sat}$ (DD2F) and $1.8\times n_\mathrm{sat}$ (DD2)}. {The numbers indicate the density jump $\Delta n$ at the phase transition in units of nuclear saturation density}. The bottom row shows hybrid models at a constant speed of sound of $c_s^2=0.7$ in the quark phase. The stiffness of the hadronic phase increases from left to right.}
\label{fig:tov}
\end{figure*}

We plot the mass-radius curves of spherically symmetric, non-rotating NSs by solving the Tolman-Oppenheimer-Volkoff (TOV) equations~\cite{Tolman:1939jz,Oppenheimer:1939ne} for a selection of generated hybrid models in Fig.~\ref{fig:tov} (colored) as well as for the underlying hadronic EoSs (black). The columns display results based on a specific hadronic model. The top row shows models at a fixed onset density with different density jumps and different sound speeds.
In the bottom row, we plot models with different density jumps and onset densities for fixed sound speed $c_s^2=0.7$.
Generally, larger density jumps lead to smaller NS radii and maximum masses, whereas larger sound speeds result in larger maximum masses and radii. For large density jumps, twin star solutions exist where the hybrid star branch is disconnected from the hadronic star branch~\cite{Gerlach:1968zz,Kampfer:1981yr,Glendenning:1998ag,Schertler:2000xq}. In total, we consider 245 different hybrid EoSs.

\subsection{Simulation code}\label{subsec:code}
We perform merger simulations using the general relativistic smoothed particle hydrodynamics (SPH) code of Refs.~\cite{Oechslin:2001km,Oechslin:2006uk,Bauswein:2009im,Blacker:2023afl}. This code solves the Einstein field equations employing the conformal flatness condition~\cite{Isenberg1980,Wilson:1996ty}. {The code employs a cubic spline kernel with an adaptive smoothing length~\cite{Monaghan1985}. A time-dependent artificial viscosity scheme is included with an additional switch to reduce viscosity in pure shear flows~\cite{Balsara1995,Chow1997,Morris1997,Oechslin2007}. } Since the hybrid models we employ (see Sect.~\ref{subsec:eos}) are purely barotropic EoSs\footnote{The hadronic models SFHo, DD2F and DD2 are available as fully temperature and composition dependent tables. However, in order to implement thermal effects of quark matter and the temperature dependence of the phase boundaries, we only employ the zero temperature beta-equilibrium slice of these hadronic tables and approximate thermal effects in the hadronic phase. This allows to use the scheme developed in Ref.~\cite{Blacker:2023afl} for modeling thermal effects in hybrid EoSs, which is necessary because the quark matter EoSs of our large sample are only available as barotropes.}, we additionally include thermal effects using the effPT scheme we presented in Ref.~\cite{Blacker:2023afl}. This scheme extends the commonly used ideal-gas approach~\cite{Janka1993,Bauswein:2010dn} by also taking into account the temperature-dependence of the hadron-quark phase transition. It assumes two separated phases of matter connected with a flat, i.e. $\frac{\mathrm{d}P}{\mathrm{d}n}=0$, coexistence region (Maxwell construction) at all temperatures. We briefly summarize the main features of this scheme as well as some slight modifications compared to Ref.~\cite{Blacker:2023afl} in Appendix~\ref{app:effpt}. 

The scheme adopts separate thermal indices $\Gamma_\mathrm{th,h}=1.75$ and $\Gamma_\mathrm{th,q}=4/3$ for the hadronic and the pure deconfined quark phase and also requires knowledge of the phase boundaries of the coexistence phase. For this, we pick a simple analytic description that remains almost constant at low temperatures below roughly 10 MeV, while featuring a moderate shift towards lower densities at several 10~MeV, i.e. the largest temperatures reached in the merger remnant. At very high temperatures round 100~MeV, it mimics the existence of a critical point. See Appendix~\ref{app:effpt} for details. 

As we demonstrated in Ref.~\cite{Blacker:2023afl}, the exact shape of the phase boundaries can have a significant impact on the dynamics and observables of NS mergers. Here, we limit ourselves to a single model of the phase boundaries, as we focus on how properties of the cold, hybrid EoS influence the dynamics and GW signal of mergers.

{The comparison in~\cite{Blacker:2023afl} shows a very good agreement between merger simulations with fully temperature and composition dependent EoS models and calculations with the effective scheme (see Tab.~II in~\cite{Blacker:2023afl} with postmerger GW frequencies coinciding to within a few 10~Hz.).}

All simulations are set up using cold, irrotational $1.35-1.35~M_\odot$ binaries at an initial center-to-center separation of around 38~km. The stars then merge within a few orbits. {Each star is modeled by about 150,000 SPH particles and the number of SPH particle neighbors is kept at about 100 throughout the simulation. For an SPH particle number in this range, the impact of resolution on the dominant GW frequency is small (a few 10~Hz,~e.g.~\cite{Bauswein2012,Kochankovski2025}). } The effects of neutrinos and magnetic fields are not included. In total, we perform 245 simulations using hybrid models, 71 based the DD2F, 106 based on the DD2 and 68 based on the SFHo EoS\footnote{The specific models and their simulation results are available in the Supplemental Material~\cite{supp}.}. 

\section{Results}\label{sec:results}

To study which combination of $n_\mathrm{on},~\Delta n$ and $c_s^2$ leaves a detectable imprint on the GW signal, we follow the ideas of Refs.~\cite{Bauswein:2018bma, Blacker:2020nlq} to identify a transition. These works showed that a tight relation existing for hadronic EoSs between the tidal deformability $\Lambda$~\cite{Flanagan:2007ix,Hinderer:2007mb,Hinderer:2009ca,Damour:2009wj,Damour:2012yf} of the inspiraling stars and the dominant postmerger GW frequency $f_\mathrm{peak}$ is violated in systems where a strong phase transition to quark matter takes place after the merger. The strong softening of the EoS increases the compactness of the remnant characteristically elevating $f_\mathrm{peak}$ (see Fig.~2 in Ref.~\cite{Bauswein:2018bma} and Fig.~4 in Ref.~\cite{Blacker:2020nlq}).

To quantify the impact and detectability of the phase transition, we define the frequency shift $\Delta f_\mathrm{peak}=f_\mathrm{peak,hybrid}-f_\mathrm{peak,had}$. $f_\mathrm{peak,hybrid}$ is the dominant postmerger GW frequency we infer from our simulation for a hybrid model. $f_\mathrm{peak,had}$ is the frequency predicted by the relation between $f_\mathrm{peak}$ and $\Lambda$ for purely hadronic EoSs. We calculate this quantity using the formula
\begin{align}
    f_\mathrm{peak,had}=a\Lambda_{1.35}^2+b\Lambda_{1.35}+c \label{eq:fpeak_lambda}
\end{align}
with $a=8.463\times 10^{-7}$~kHz, $b=−2.509\times 10^{-3}$~kHz and $c=4.182$~kHz from Ref.~\cite{Blacker:2020nlq} using the tidal deformability of a 1.35~$M_\odot$ star from the respective EoS of the inspiraling stars \footnote{Note that we will also refer to this value as $f_\mathrm{peak,had}$ if the $n_\mathrm{on}$ is below the central density of a $1.35~M_\odot$ star, i.e. when the inspiraling stars are hybrid stars. { $f_\mathrm{peak,had}$ is the fiducial value one expects for a given tidal deformability if there is no quark matter present at all.}}. 

The frequency shift $\Delta f_\mathrm{peak}$ hence quantifies the difference between the actual postmerger frequency and the expected $f_\mathrm{peak}$ based on the tidal deformability of a 1.35~$M_\odot$ star assuming the underlying EoSs is purely hadronic. A sizable frequency shift would provide strong evidence for the occurrence of deconfined quark matter.

\begin{figure} 
\centering
\includegraphics[width=1.0\linewidth]{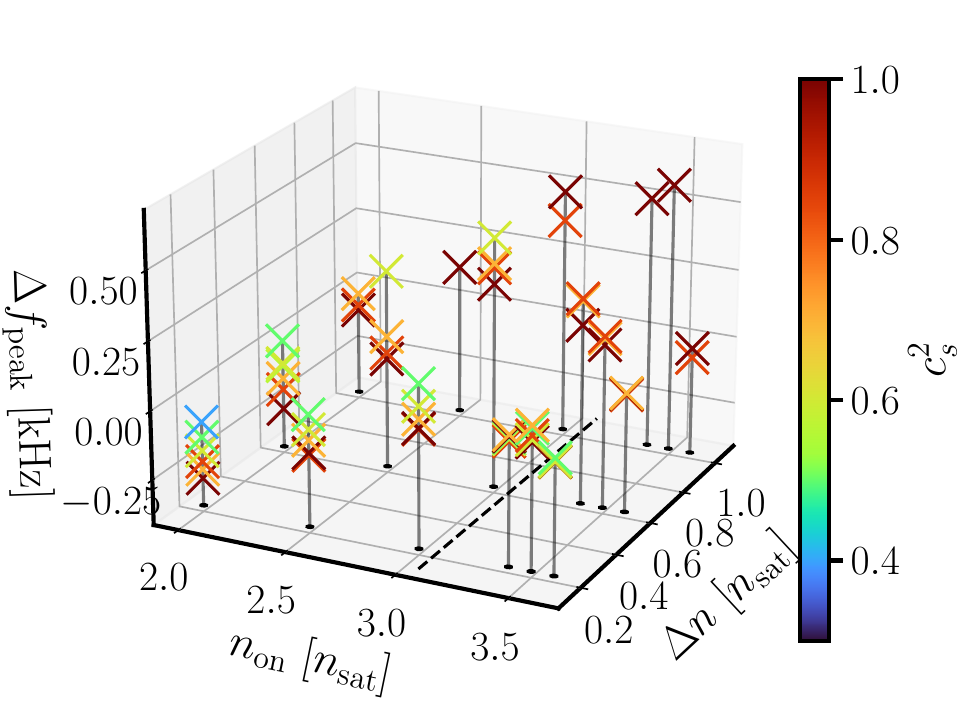}
\caption{Shift of the dominant postmerger gravitational-wave frequency $f_\mathrm{peak}$ in merger simulations using SFHo-based hybrid EoSs compared to $f_\mathrm{peak}$ predicted by Eq.~\eqref{eq:fpeak_lambda} based on the tidal deformability of the inspiraling stars (i.e. assuming a hadronic EoS) as a function of the onset density, density jump and stiffness of the quark phase. The dashed line marks the central density of the inspiraling purely hadronic stars, i.e. for SFHo. { Hence, models at smaller values of $n_\mathrm{on}$ contain quark matter before merging, whereas models to the right of the dashed line are purely hadronic during the inspiral.}}
\label{fig:df_DD2F}
\end{figure}

\begin{figure} 
\centering
\includegraphics[width=1.0\linewidth]{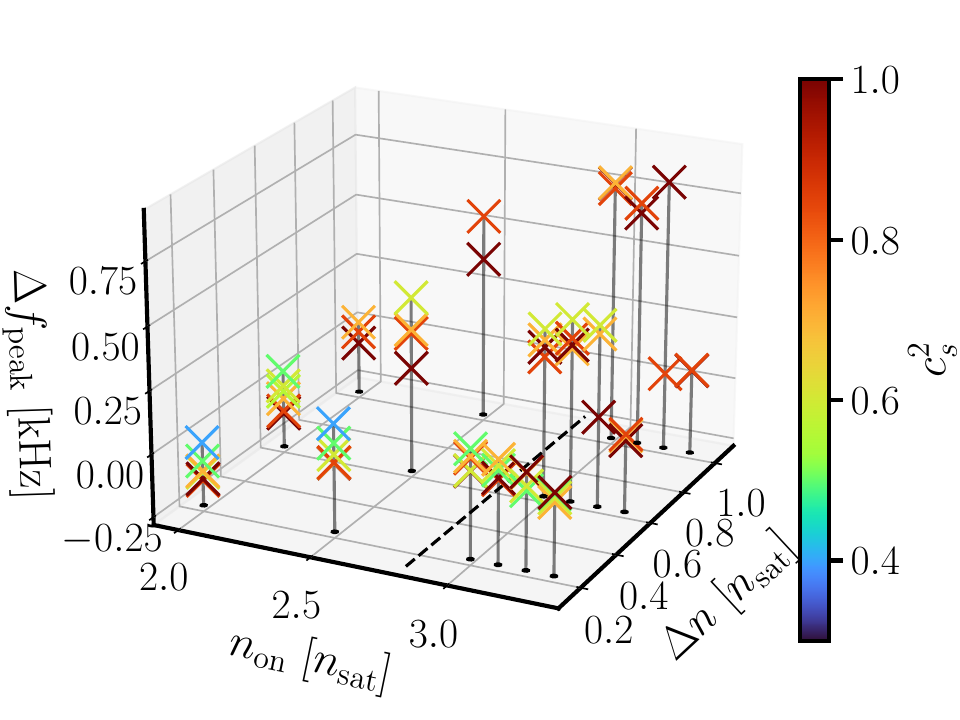}
\caption{Same as Fig.~\ref{fig:df_DD2F}, but for DD2F-based models.}
\label{fig:df_DD2}
\end{figure}

\begin{figure} 
\centering
\includegraphics[width=1.0\linewidth]{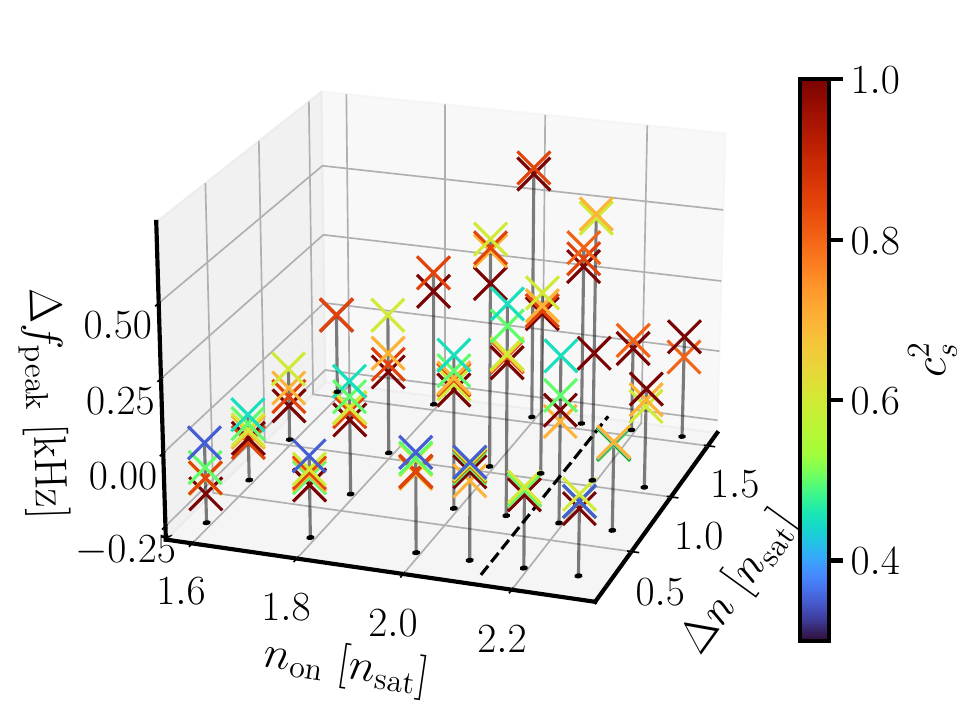}
\caption{Same as Fig.~\ref{fig:df_DD2F}, but for DD2-based models.}
\label{fig:df_SFHO}
\end{figure}

In Figs.~\ref{fig:df_DD2F},~\ref{fig:df_DD2} and~\ref{fig:df_SFHO} we plot $\Delta f_\mathrm{peak}$ as a function of $n_\mathrm{on}$ and $\Delta n$ for all our simulations that do not result in a prompt collapse at merger (see Appendix~\ref{app:mub} for plots using the baryon chemical potential { showing similar functional dependencies}). The coloring indicates $c_s^2$ in the quark phase. The dashed lines highlight the central density of the inspiraling stars. This means that in all systems with smaller $n_\mathrm{on}$ deconfined quark matter is already present prior to the merger while in all systems with larger $n_\mathrm{on}$ the two NSs are purely hadronic. We find that $\Delta f_\mathrm{peak}$ can vary significantly up to almost 750~Hz depending on the parameters of the hybrid EoS. Generally, $\Delta f_\mathrm{peak}$ grows with $\Delta n$ and decreases more weakly with $c_s^2$. This is understandable, since a larger $\Delta n$ leads to a more compact remnant while a stiffer quark phase  reduces the compactness of the quark core compared to a softer quark phase. For very large $\Delta n$ a prompt collapse occurs (no data point shown).

The dependence of $\Delta f_\mathrm{peak}$ on the onset density of the phase transition is more complex. At low $n_\mathrm{on}$, we find $\Delta f_\mathrm{peak}\approx 0$ followed by an increase in $\Delta f_\mathrm{peak}$ with
$n_\mathrm{on}$ eventually reaching a maximum. A further increase in $n_\mathrm{on}$ results in decreasing frequency shifts. This behavior is caused by counteracting effects. On the one hand, at low onset densities the inspiraling stars are hybrid stars and the tidal deformability is also affected by the phase transition increasing $f_\mathrm{peak,had}$. Hence, the transition is not visible by comparing $\Lambda$ and $f_\mathrm{peak}$. Additionally, the quark phase is typically stiffer than the hadronic phase, as can be seen in Fig.~\ref{fig:tov} by the larger maximum masses of hybrid EoSs with small density jumps compared to the purely hadronic EoSs. A low onset density, i.e. a large quark core therefore results in relatively low remnant densities despite the softening from the phase transition. {For some models with low $n_\mathrm{on}$, we find a negative frequency shift, i.e. the hybrid model with quark matter leads to a decrease of the dominant postmerger GW frequency compared to the one from the corresponding purely nucleonic model (cf.~\cite{2024arXiv240709446H}).}

An increase in $n_\mathrm{on}$ leads to a smaller quark core, but also to higher densities and hence first to an overall more compact remnant. In addition, hybrid EoSs with less quark matter have a larger tidal deformability hence decreasing $f_\mathrm{peak,had}$. Both effects result in larger $\Delta f_\mathrm{peak}$.

On the other hand, the size of the soft phase transition region in the remnant shrinks for larger $n_\mathrm{on}$ hence reducing $\Delta f_\mathrm{peak}$ and making the overall remnant structure more similar to a purely hadronic scenario. Both effects together lead to the local maximum in $\Delta f_\mathrm{peak}$ for varying $n_\mathrm{on}$ and fixed $\Delta n$ and $c_s^2$. We find that the decrease in $\Delta f_\mathrm{peak}$ at high $n_\mathrm{on}$ can occur rapidly over a narrow $n_\mathrm{on}$ range.

We remark that the magnitude of the frequency shift $\Delta f_\mathrm{peak}$ of the hybrid star mergers in Ref.~\cite{Bauswein:2020ggy} somewhat differ from our values at similar $n_\mathrm{on}=1.8\times n_\mathrm{sat}$ and $\Delta n=1.45\times n_\mathrm{sat}$ with the DD2 base hadronic model. These discrepancies can be attributed to the different finite-temperature behavior of the phase boundary {adopted in this work as compared to~\cite{Bauswein:2020ggy}.} The transition region in the models employed by Ref.~\cite{Bauswein:2020ggy} is shifted stronger towards lower densities with increasing temperature than in this study (compare Fig.~\ref{fig:bounds} in the appendix and Fig.~9 in Ref.~\cite{Bastian:2020unt}).

For DD2- and DD2F-based models, we observe two peaks in the GW spectra at $n_\mathrm{on}=2.3\times n_\mathrm{sat}$ and $n_\mathrm{on}=3.2\times n_\mathrm{sat}$, respectively for $\Delta n \geq 0.6 \times n_\mathrm{sat}$. Such a feature was also reported in Ref.~\cite{Weih:2019xvw} and is indicative of a delayed transition in the merger remnant. Hence, the GW spectrum shows a peak at smaller frequencies from the early remnant evolution phase and a second peak at higher frequencies once a sizable quark core has formed. In Figs.~\ref{fig:df_DD2F},~\ref{fig:df_DD2} and~\ref{fig:df_SFHO}, we always plot the strongest peak in the GW spectrum, which in the case of a two-peaked spectrum sometimes corresponds to the peak at lower and sometimes at higher frequency. This explains the large differences in $\Delta f_\mathrm{peak}$ visible at $n_\mathrm{on}=3.2\times n_\mathrm{sat}$ in Fig.~\ref{fig:df_DD2F} and at $n_\mathrm{on}=2.3\times n_\mathrm{sat}$ in Fig.~\ref{fig:df_DD2} for fixed $\Delta n$. These double peaks, however, occur only in a very narrow region of the parameter space.

At higher $n_\mathrm{on}$, the quark matter core becomes too small and the phase transition is no longer visible, i.e. $\Delta f_\mathrm{peak} \approx 0$. For SFHo-based hybrid EoSs, we do not observe a double peak structure for any combination of parameters and the phase transition leaves no imprint on the GW signal for $n_\mathrm{on}\geq 3.6\times n_\mathrm{sat}$.

For large density jumps we observe prompt black-hole formation at merger for all  $c_s^2$ of the quark phase. We have highlighted these configurations with plus signs in Fig.~\ref{fig:c2_constraints}.
\begin{figure*}
\centering
\subfigure[]{\includegraphics[width=0.33\linewidth]{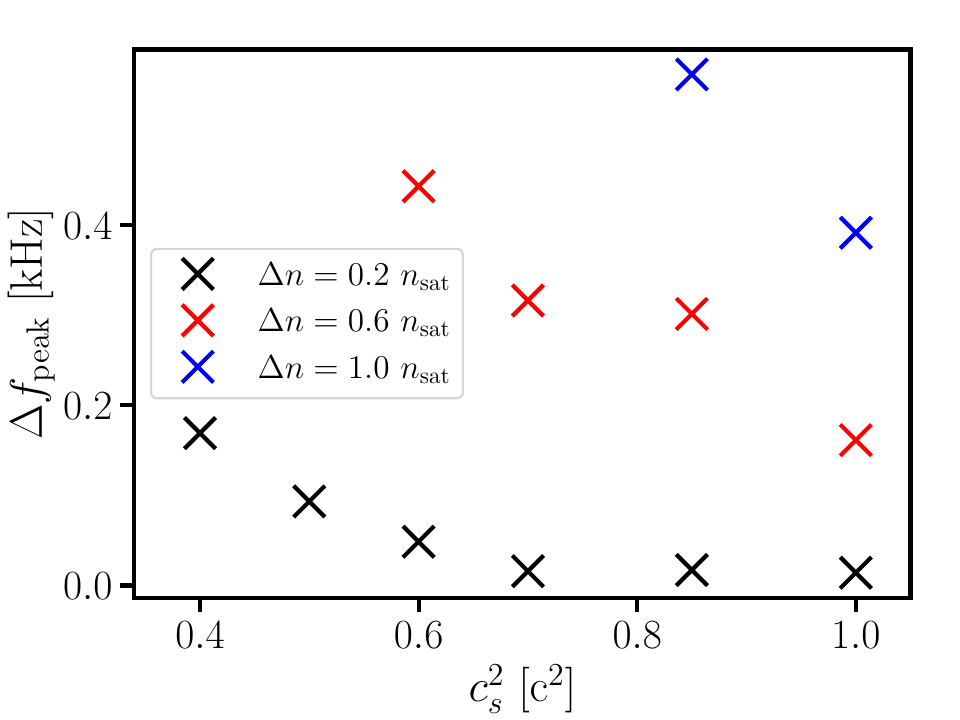}\label{fig:1dresult_a}}
\hfill
\subfigure[]{\includegraphics[width=0.32\linewidth]{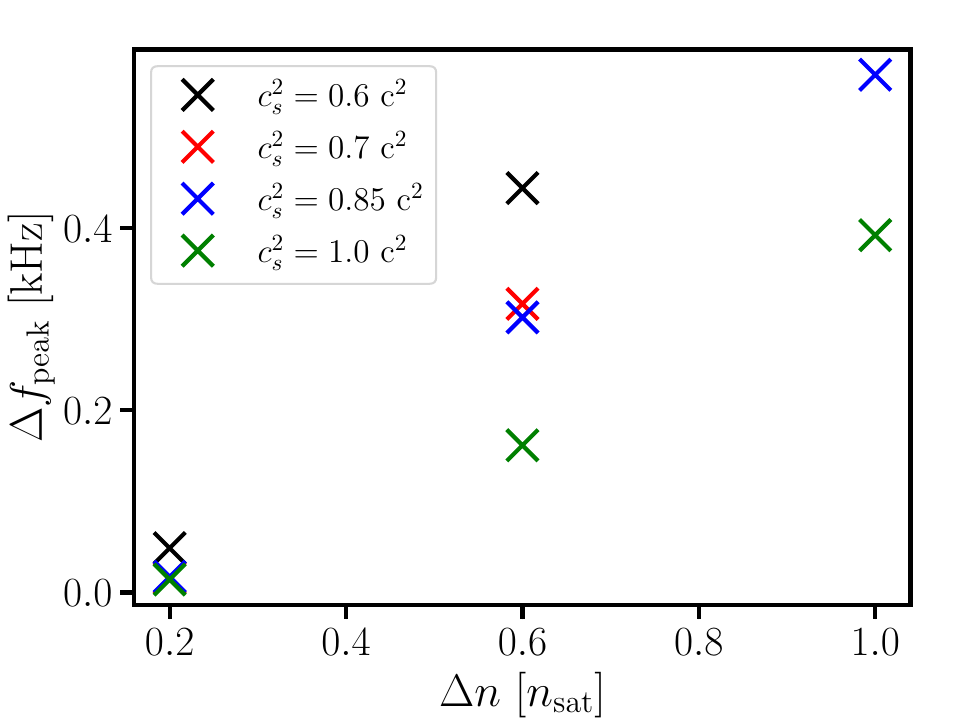}\label{fig:1dresult_b}}
\subfigure[]{\includegraphics[width=0.32\linewidth]{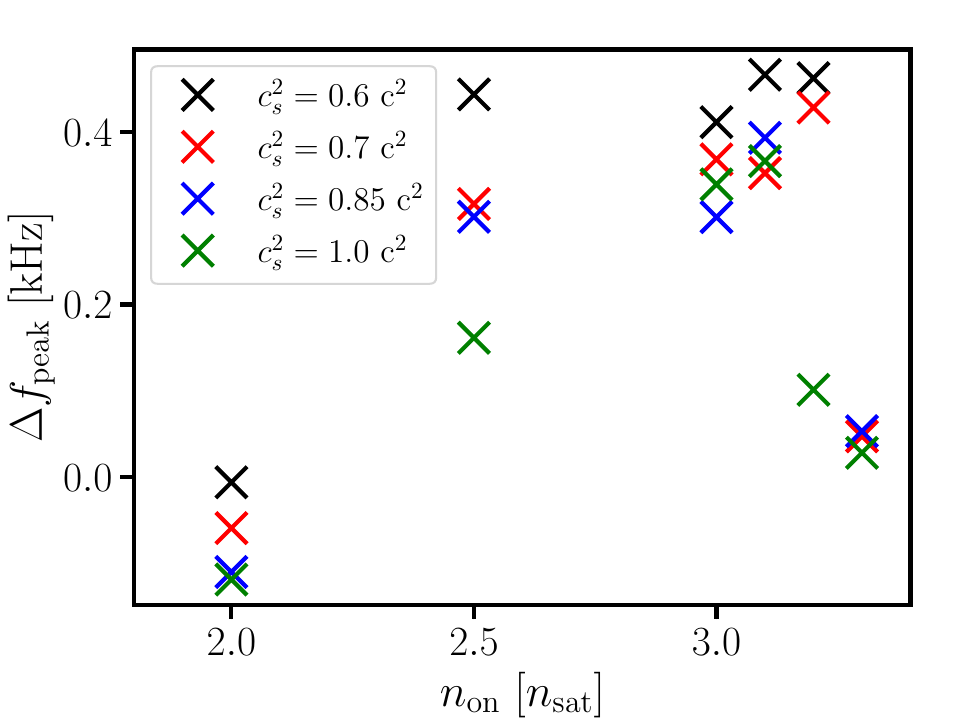}\label{fig:1dresult_c}}
\hfill
\caption{Shift of the dominant postmerger gravitational wave frequency $\Delta f_\mathrm{peak}$ in merger simulations using DD2F-based hybrid EoSs compared to $f_\mathrm{peak}$ predicted by Eq.~\eqref{eq:fpeak_lambda} based on the tidal deformability of the inspiraling stars as a function of different hybrid EoS properties. (a): $\Delta f_\mathrm{peak}$ as a function of $c_s^2$ for $n_\mathrm{on}=2.5 \times n_\mathrm{sat}$. Different colors refer to different $\Delta n$ at the phase transition. (b): $\Delta f_\mathrm{peak}$ as a function of $\Delta n$ for $n_\mathrm{on}=2.5 \times n_\mathrm{sat}$. (c): $\Delta f_\mathrm{peak}$ as a function of $n_\mathrm{on}$ for $\Delta n=0.6 \times n_\mathrm{sat}$. In the latter two panels different colors refer to results from models with different $c^2_s$.}
\label{fig:1dresult}
\end{figure*}

In a next step, we quantify how $\Delta f_\mathrm{peak}$ depends on the properties of the hybrid EoS by providing an empirical formula we fit to our data.
To motivate the form of this formula, we visualize the dependence of $\Delta f_\mathrm{peak}$ on a single parameter $n_\mathrm{on}$, $\Delta n$ or $c_s^2$ for DD2F-based models in Fig.~\ref{fig:1dresult}. In each of the three panels, we keep two parameters fixed and show how $\Delta f_\mathrm{peak}$ changes with the respective third quantity. Plots for DD2- and SFHo-based models look qualitatively similar. We find that the dependence of $\Delta f_\mathrm{peak}$ on $\Delta n$ and $c_s^2$ can be well approximated by a linear relation. 

For the dependence of $\Delta f$ on $c_s^2$, a small curvature is visible in Fig.~\ref{fig:1dresult_a}. However, we still find reasonable agreement with a linear model {(better than 100~Hz)}. Since $\Delta f_\mathrm{peak}$ shows a local maximum with respect to $n_\mathrm{on}$, we employ a quadratic dependence in our empirical formula. We find that this approach describes the increase of $\Delta f$ with rising $n_\mathrm{on}$ with good accuracy, the generally faster fall-off after the maximum is however not captured so well.

We refrain from exploring more complicated fit formulae that { may} possibly provide better fits because our data set is still somewhat limited with only three hadronic base models, a simple quark matter parametrization and a single prescribed temperature dependence of the phase boundaries.

We choose an ansatz to express the changes in GW frequency caused by the phase transition for a fixed hadronic EoS. It depends on the intrinsic properties of the phase transition and reads

\begin{align}
    \Delta f_\mathrm{peak}=(a\Delta n+b)n_\mathrm{on}^2+c\Delta n n_\mathrm{on}+d\Delta n +e c^2_s~. \label{eq:fitall}
\end{align}
We fit this formula with $a,~b,~c,~d$ and $e$ being fit parameters to our dataset for each hadronic model separately using a least squares fit. This is motivated by the expectation that by the time $f_\mathrm{peak}$ will be measured, the EoS at lower density and NS mass will be relatively well known e.g. from GW inspiral detections. We omit to provide a general formula independent of the hadronic EoS since the range of $n_\mathrm{on}$ where $\Delta f_\mathrm{peak}$ is substantially larger than zero depends strongly on the properties of hadronic matter. We restrict the fits to the onset density ranges $n_\mathrm{on} \in [2.0\times n_\mathrm{sat},~3.1\times n_\mathrm{sat}]$ for DD2F-based models, $n_\mathrm{on} \in [1.6\times n_\mathrm{sat},~2.1\times n_\mathrm{sat}]$ for DD2-based models and $n_\mathrm{on} \in [2.0\times n_\mathrm{sat},~3.5\times n_\mathrm{sat}]$ for SFHo-based models, as in this region the amount of quark matter in the remnant is sufficient to influence $f_\mathrm{peak}$ and there is a unique dominant postmerger GW frequency visible in the spectrum. At larger $n_\mathrm{on}$, $\Delta f_\mathrm{peak}$ drops to zero because no or only small amounts of quark matter are present in the merger remnant. We provide an additional argument for these ranges in Appendix.~\ref{app:ranges}.

The fit parameters together with the mean and maximum deviation of the model from the respective data are provided in Tab.~\ref{tab:fits}.

\begin{table*}
\caption{Fit parameters of Eq.~\eqref{eq:fitall} for our three sets of hybrid EoSs as well as the mean and the maximum deviation of our data from the fit. The last two columns provide the fit range.}
\begin{tabular}{c | c c c c c c c c c}
\hline\hline
nuc. & $a$ & $b$ & $c$ & $d$ & $e$ & max. & mean & min & max \\
 & & & & & & dev. & dev. & $n_\mathrm{on}$ & $n_\mathrm{on}$ \\
EoSs &  &  &  &  &  & $\left[ \mathrm{kHz}\right] $ & $\left[ \mathrm{kHz}\right] $ & $\left[ n_\mathrm{sat}\right] $ & $\left[ n_\mathrm{sat}\right] $ \\
\hline\
DD2F & -0.752 & 0.013 & 4.439 & -5.730 & -0.249 & 0.111 & 0.035 & 2.0 & 3.1 \\
DD2 & -1.308 & 0.028 & 5.588 & -5.571 & -0.234 & 0.203 & 0.062  & 1.6 & 2.1 \\
SFHo & -0.445 & 0.018 & 2.796 & -3.684 & -0.362 & 0.156 & 0.059 & 2.0 & 3.5 \\
\hline
\hline
\end{tabular}
\label{tab:fits}
\end{table*}

In general, this relation captures the overall behavior of the data relatively well with a mean scatter up to around 60~Hz. However, we also observe deviations from the fit up to 200~Hz, for DD2-based models. The relation we found is hence not very tight for all data points although on average it permits a decent approximation. 

The relation in Eq.~\eqref{eq:fitall} can be used in the future to constrain the properties of a phase transition based on an observed $f_\mathrm{peak}$, once sufficiently precise GW measurements are available~\cite{Chatziioannou:2017ixj,Wijngaarden:2022sah}. {In this regard we mention that the postmerger signal-to-noise ratios (SNRs) of GW signals from simulations with the hybrid EoSs are very comparable to those of purely nucleonic models (for noise curves of a single Advanced LIGO detector at design sensitivity~\cite{ligodesign} the SNRs vary between 0.35 and 2.17 with an average of 1.36 across all hybrid models compared to 1.5295, 1.497, and 2.0205 for the nucleonic EoSs SFHO, DD2F and DD2 at a distance of 40~Mpc along the polar direction\footnote{Note that the SNR increases if a network of detectors is used.}). For a signal comparably loud as that of GW170817, an improvement of the sensitivity by factors of a few compared to Advanced LIGO's design sensitivity is required to extract $f_\mathrm{peak}$~\cite{Chatziioannou:2017ixj,Torres2019,Wijngaarden:2022sah}.} First, $f_\mathrm{peak}$ together with the inferred tidal deformability can be compared to \eqref{eq:fpeak_lambda} (or an equivalent formula if the binary mass differs significantly from $2\times 1.35~\mathrm{M}_\odot$; {see e.g.~\cite{Blacker:2020nlq} for typical variations with $M_\mathrm{tot}$ and $q$, which are for instance of the order of a few 10~Hz for changes of $10^{-2}~M_\odot$ in $M_\mathrm{tot}$}) to obtain $f_\mathrm{peak,had}$ and hence $\Delta f_\mathrm{peak}$. If $\Delta f_\mathrm{peak}$ is at least $\sim$200~Hz, this indicates the presence of a phase transition in the merger remnant. Then, Eq.~\eqref{eq:fitall} can be inverted to obtain constraints on $c_s$, $n_\mathrm{on}$ or $\Delta n$. {Employing a refined fit in a smaller parameter range, comparing directly to the data, and computing additional models in a preselected range may be advantageous in this respect.}

\begin{figure*} 
\centering
\subfigure[]{\includegraphics[width=0.33\linewidth]{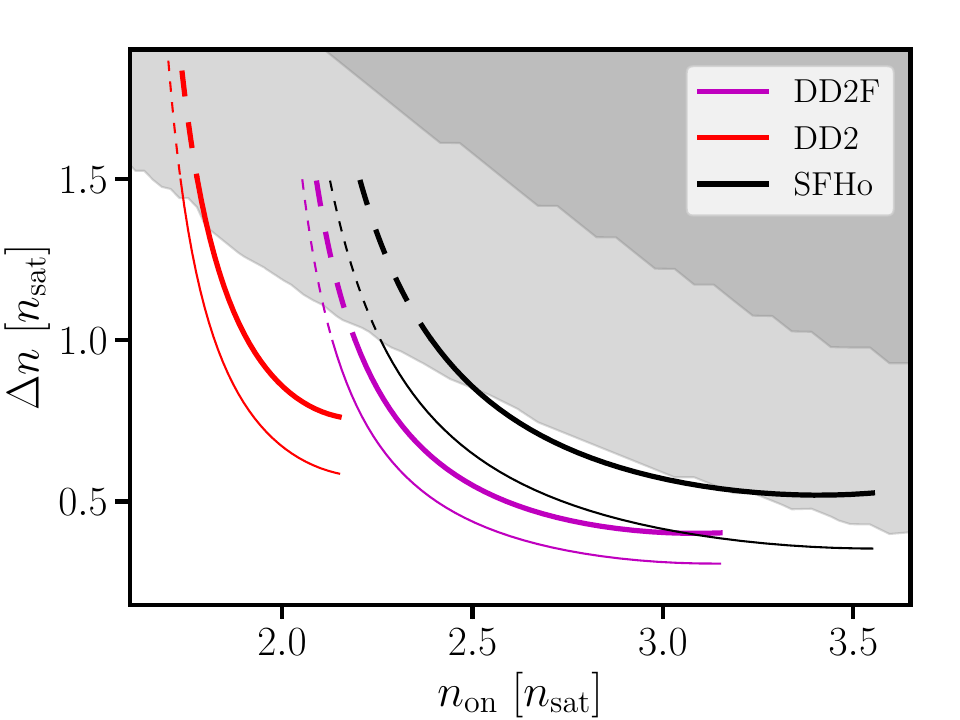}\label{fig:7dresult_a}}
\hfill
\subfigure[]{\includegraphics[width=0.32\linewidth]{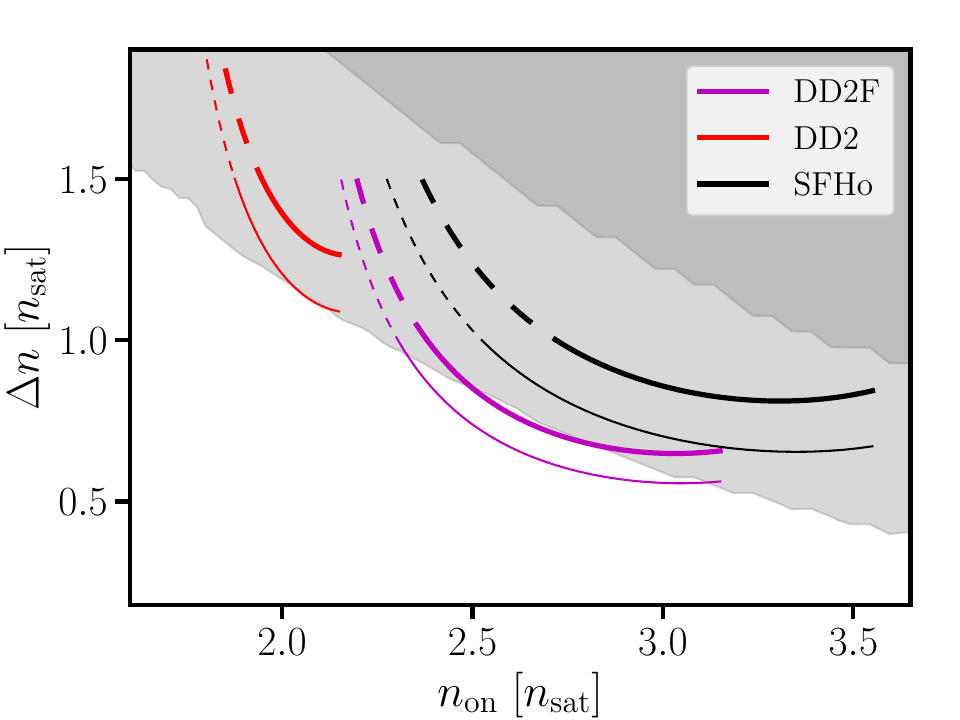}\label{fig:7dresult_b}}
\subfigure[]{\includegraphics[width=0.32\linewidth]{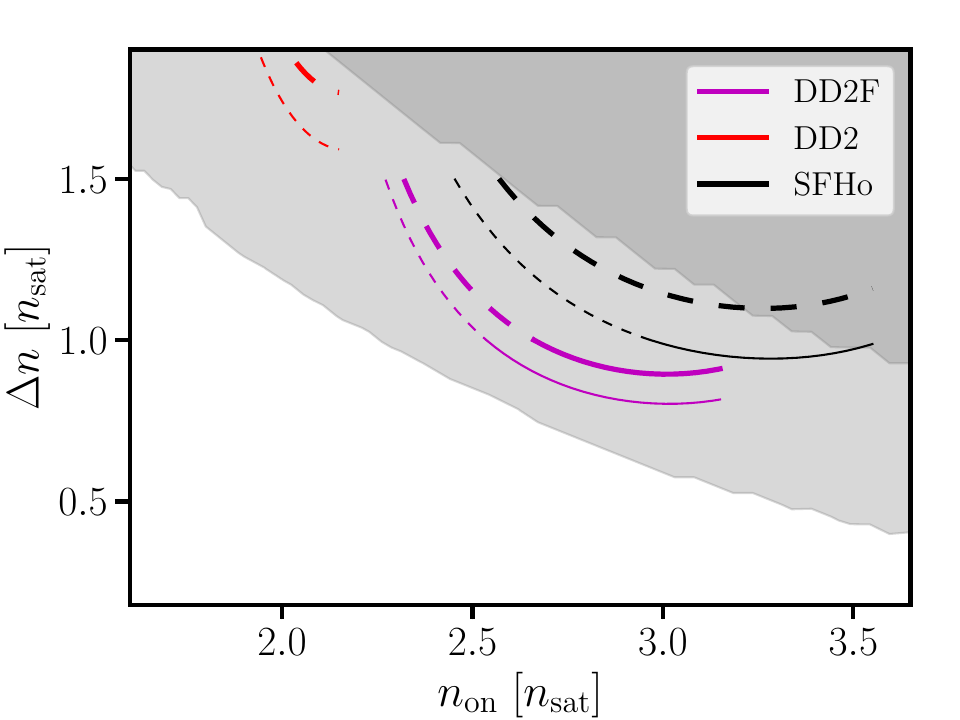}\label{fig:7dresult_c}}
\hfill
\caption{Combinations of density jump $\Delta n$ and onset density $n_\mathrm{on}$ for a fixed frequency shift of $\Delta f_\mathrm{peak}$=200~Hz (left), $\Delta f_\mathrm{peak}$=400~Hz (middle) and $\Delta f_\mathrm{peak}$=600~Hz (right) from inverting Eq.~\eqref{eq:fitall}. Different colors represent different underlying hadronic EoSs. Thick and thin lines mark curves with $c^2_s=1.0$ and $c^2_s=0.7$, respectively. Dashed lines represent configurations that are prone to collapse to a black hole. Specifically, the left end of the dashed line represents a configuration which undergoes a prompt collapse, whereas the beginning of the dashed line at the right marks an EoS model which yields a temporarily stable merger remnant. Beyond the right end of the solid lines $n_\mathrm{on}$ is so high that at most small amounts of quark matter are found in the remnant and thus $\Delta f_\mathrm{peak}$ is close to zero. The light gray-shaded area marks the region where $c_s^2\geq 0.7$ is required to fulfill the $2~M_\odot$ constraint and the darker shaded region is excluded by causality (compare Fig.~\ref{fig:c2_constraints}). Note that some curves for $c_s=0.7$ are excluded by pulsar mass measurements.}
\label{fig:7dresult}
\end{figure*}

As an example, we plot the density jump $\Delta n$ as a function of $n_\mathrm{on}$ in Fig.~\ref{fig:7dresult} assuming $\Delta f_\mathrm{peak}$=200~Hz, $\Delta f_\mathrm{peak}$=400~Hz and $\Delta f_\mathrm{peak}$=600~Hz in the different panels, respectively. Naturally, a larger $\Delta f_\mathrm{peak}$ implies a larger density jump while a smaller $\Delta f_\mathrm{peak}$ points to smaller $\Delta n$. Thick and thin lines refer to $c_s^2=0.7$ and $c_s^2=1.0$, respectively. The light gray shaded area marks the region where $c_s^2\geq 0.7$  is required to fulfill the $2~M_\odot$ constraint and the darker shaded region is excluded by causality (compare Fig.~\ref{fig:c2_constraints}). The different colors highlight the fits for hybrid EoSs based on the three different hadronic EoSs we employ.

All curves rise very steeply towards large $\Delta n$ at low $n_\mathrm{on}$. In this $n_\mathrm{on}$ range deconfined quark matter is already present in the inspiraling stars and $\Delta n$ significantly impacts both $f_\mathrm{peak}$ and $\Lambda$. Here, very large density jumps are required to make the phase transition detectable. At larger $n_\mathrm{on}$, $\Lambda$ is no longer affected and the required $\Delta n$ only weakly depends on $n_\mathrm{on}$.

\begin{figure} 
\centering
\includegraphics[width=1.0\linewidth]{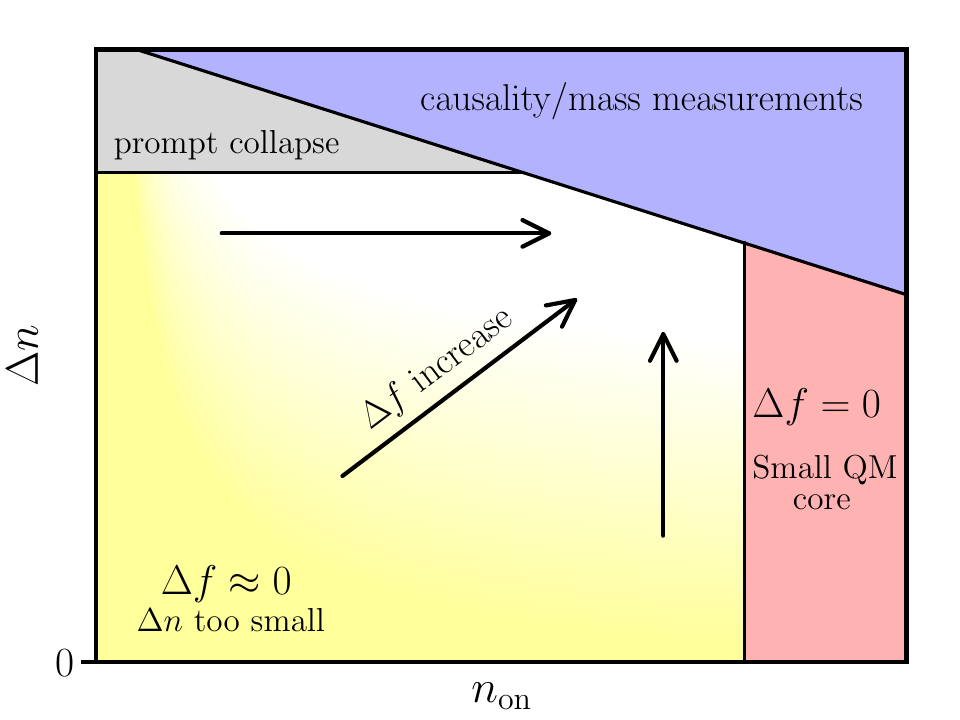}
\caption{Sketch of how the frequency shift $\Delta f_\mathrm{peak}$ depends on the onset density $n_\mathrm{on}$ and the density jump $\Delta n$ of the phase transition. Generally, $\Delta f_\mathrm{peak}$ rises with $\Delta n$ and $n_\mathrm{on}$ as indicated by the arrows. The frequency shift is small at low $\Delta n$ and/or $n_\mathrm{on}$ (yellow region). Very large values of $\Delta n$ either result in prompt black hole formation at merger (gray region) or are excluded by pulsar mass measurements (blue region). At high $n_\mathrm{on}$, the amount of quark matter in the remnant becomes too small to affect $f_\mathrm{peak}$ (red region).}
\label{fig:result_sketch}
\end{figure}

Generally, small $\Delta f_\mathrm{peak}$ are only compatible with the parametrizations towards the lower left region of the panels, while large $\Delta f_\mathrm{peak}$ indicate combinations of parameters towards the upper right region of the panels. We provide an overview sketch of the dependencies in Fig.~\ref{fig:result_sketch}.

We also emphasize that the allowed parameter space after an $f_\mathrm{peak}$ detection is limited at larger $\Delta n$ by the occurrence of a prompt collapse and by NS maximum mass and causality constraints (see Fig.~\ref{fig:c2_constraints}), as indicated by the shaded areas. At lower $\Delta n$ and relatively large $n_\mathrm{on}$ the curves terminate because larger $n_\mathrm{on}$ would prevent the formation of significant amounts of quark matter in the remnant thus yielding $\Delta f_\mathrm{peak} \approx 0$. $\Delta f_\mathrm{peak} \approx 0$ (e.g. no clear evidence of a phase transition), is only compatible with narrow regions either along the $\Delta n$-axis in a small range of $n_\mathrm{on}$, with very low $\Delta n$ or with very large $n_\mathrm{on}$ such that the quark matter amount in the remnant are small or zero.

Our formula can be used to estimate which combination of transition properties can lead to a clear difference in $f_\mathrm{peak}$ compared to a system with no phase transition. The different panels of Fig.~\ref{fig:7dresult} give a visual impression on the magnitude of $\Delta f_\mathrm{peak}$ that can be expected to be sizable roughly in the range between $2\leq n_\mathrm{sat}\leq 3.5$ and $0.5\leq \Delta n\leq 1.5$ (see also Figs.~\ref{fig:df_DD2F}-\ref{fig:df_SFHO}). Note that for $c_s$ significantly below 0.7 the parameter range which leads to sizable $\Delta f_\mathrm{peak}$ shrinks further (not shown in Fig.~\ref{fig:7dresult} but visible in Figs.~\ref{fig:df_DD2F},~\ref{fig:df_DD2} and~\ref{fig:df_SFHO}). Curves for smaller $c_s$ lie below those shown in Fig.~\ref{fig:7dresult} but are stronger constrained by pulsar mass measurements. This, in turn, implies constraints on $c_s$, e.g., larger $\Delta f_\mathrm{peak}$ are more compatible with high quark matter stiffness (despite the opposite trend being indicated in the left panel of Fig.~\ref{fig:1dresult}, which is superseded by the impact of $\Delta n$).

Hypothetically assuming that a future detection would yield a finite $\Delta f_\mathrm{peak}$, this would constrain the allowed values of $n_\mathrm{on}$ and $\Delta n$ to a relatively narrow range depending on the hadronic EoS below $n_\mathrm{on}$ {(see Figs.~\ref{fig:df_DD2F},~\ref{fig:df_DD2} and~\ref{fig:df_SFHO} for the range of phase transition properties which yield a finite $\Delta f_\mathrm{peak}$).}. We remark that once sufficiently precise GW measurements to infer $f_\mathrm{peak}$ are available, the cold EoS at lower densities may be relatively well known from GW inspiral signals, but we also refer to the limitations of our current study briefly addressed in the next section, which may imply a moderate broadening of the parameter ranges discussed above, e.g., by other hadronic base models, different phase boundaries at finite temperature or other quark matter models.

In Appendix~\ref{app:mub} we provide figures equivalent to Fig.~\ref{fig:7dresult} but replacing the onset density by the baryon chemical potential at the phase transition.

\section{Summary and discussion}\label{sec:summary}
In this work, we perform a large number of simulations with a constant speed of sound parametrization for the quark phase and three different underlying hadronic models. We systematically vary the stiffness of the quark phase as well as the onset density and the density jump of the phase transition to explore how properties of the hadron-quark phase transition affect the postmerger GW signal of NS mergers.

We discuss the shift in $f_\mathrm{peak}$ caused by the phase transition relative to the frequency one would expect if the EoS was purely hadronic. We find that this shift $\Delta f_\mathrm{peak}$ in the dominant postmerger gravitational-wave frequency compared to the empirical relation of Ref.~\cite{Blacker:2020nlq} for purely hadronic EoS scales approximately linear with $c_\mathrm{s}^{2}$ and $\Delta n$, where changes in $\Delta n$ affect $f_\mathrm{peak}$ stronger than different $c_\mathrm{s}^{2}$. Small $\Delta n$ typically lead to small shifts in $f_\mathrm{peak}$ whereas at large $\Delta n$ the inference of $\Delta f_\mathrm{peak}$ is limited by the occurrence of prompt black hole formation.

Regarding the dependence on $n_\mathrm{on}$, we observe that $\Delta f_\mathrm{peak}$ exhibits a maximum at intermediate values. At low and high $n_\mathrm{on}$, we find $\Delta f_\mathrm{peak}\approx 0$. A clear detection of a phase transition through a finite $\Delta f_\mathrm{peak}$ hence significantly constrains the parameter range. For a given finite value of  $\Delta f_\mathrm{peak}$ the allowed properties of the hadron-quark phase transition become narrowly constrained. 

From our data, we derive an empirical formula for the shift by quantifying $\Delta f_\mathrm{peak}$ as a function of $n_\mathrm{on}$, $c_\mathrm{s}^{2}$ and $\Delta n$ for each set of models with a fixed hadronic part. We demonstrate how this relation can be used to constrain properties of the hadron-quark phase transition if future GW detections and EoS constraints indicate the presence of such a transition in NS merger remnants. If no clear signs of a transition are present, our relation can also be used to exclude regions of the parameter space and to identify the still viable properties of the transition to deconfined quark matter. This can be complemented by constraints from current heavy-ion experiments such as HADES~\cite{HADES:2019auv} or future experimental facilities like FAIR~\cite{Friman:2011zz,Senger:2021tlg} and NICA~\cite{Blaschke2016,MPD:2022qhn}. {Such complementary information is in particular valuable for small or negligible frequency shifts, which can either imply purely baryonic matter or a transition to quark matter that masquerades baryonic matter~\cite{Alford:2004pf}. We also note that small frequency shifts of the order of 100~Hz {can be caused by specific thermal properties of nucleonic matter~\cite{Raithel2023,Fields2023,Raithel2025} or} can point to the occurrence of hyperonic degrees of freedom~\cite{Blacker2024,Kochankovski2025}.}

We conclude with a couple of remarks explaining the limitations our current study and the necessity of future work, which, however, can be basically conducted within the framework presented here. In this work we, have assumed a single prescription of the finite temperature phase boundaries. As we have demonstrated in Ref~\cite{Blacker:2023afl}, the finite temperature behavior of the hadron-quark phase transition can have a significant impact on the dynamics of the merger and hence on the postmerger GW signal. For models with a stronger shift of the transition region to lower densities, we expect a noticeable impact on the GW signal for a larger $n_\mathrm{on}$ range than in this work. Specifically, at low densities the transition may still be visible despite the transition also affecting $\Lambda$, as in Ref.~\cite{Bauswein:2020ggy}. An important extension of our work would therefore be to vary the hot EoSs, specifically the temperature dependence of the phase boundaries. 

We have only considered a transitions with a Maxwell construction, a single parametric quark model and three underlying hadronic EoSs. Other types of transitions allowing for coexisting charged phases~\cite{Glendenning:1992vb,Hempel:2013tfa,Hammond2025} or a smooth crossover~\cite{Baym:2017whm,HotQCD:2018pds} are possible and should also be thoroughly investigated {as well as the role of non-convexity of the EoS and magnetic fields~\cite{Aloy:2018jov, Rivieccio:2024sfm,Tsokaros2025}}. It is also important to employ more sophisticated microphysical quark matter models and more hadronic EoSs in future work. 

We have limited ourselves to a fixed binary mass. While we do not expect large qualitative differences for other binary masses, quantitative changes of $\Delta f_\mathrm{peak}$ with varying total mass and mass ratio will appear. For instance, it is conceivable that the remnant reaches higher densities for larger binary masses and thus quark matter may affect $f_\mathrm{peak}$ even for larger $n_\mathrm{on}$ (see~\cite{Blacker:2020nlq}). Since the threshold of prompt black hole formation depends sensitively on the properties of the EoSs, a similarly extensive exploration of this threshold in the context of hybrid EoSs will be insightful~\cite{Bauswein:2020aag,Bauswein:2020xlt,Ecker:2024kzs}.

\acknowledgements{This work was funded by Deutsche Forschungsgemeinschaft (DFG, German Research Foundation) - Project-ID 279384907 - SFB 1245. AB acknowledges support by the European Research Council under the European Union’s Horizon 2020 research and innovation programme under Grant No. 101071865 (ERC Grant HEAVYMETAL) and by the State of Hesse within the Cluster Project ELEMENTS.}

%

\begin{appendix}

\section{Hybrid EoS model}\label{app:eos}
We construct different hybrid EoSs by employing a constant speed of sound parametrization of the quark phase by following the scheme of Ref.~\cite{Chamel:2012ea,Zdunik:2012dj,Alford:2013aca}. In this approach, the pressure $P$ at zero temperature is given by

\begin{align}
    P=A\left(\frac{\mu_B}{\mu_x}\right)^{1+\beta}-B~,\label{eq:pQ}
\end{align}
where $\mu_B$ is the baryon chemical potential. $A$, $B$ and $\beta$ are free parameters where $\beta$ is related to the squared sound speed of the quark phase $c_s^2$ via 
\begin{align}
    c_s^2=\frac{\mathrm{d}P}{\mathrm{d}e}=1/\beta~.
\end{align}
$\mu_x$ is a scaling variable for the chemical potential, which we set to $\mu_x=$ 1~GeV.
The baryon number density $n_B$ and the energy density $e$ are then given by
\begin{align}
    n_B&=\frac{\mathrm{d}P}{\mathrm{d}\mu_B}=A\frac{1+\beta}{\mu_x}\left(\frac{\mu_B}{\mu_x}\right)^\beta\\
    e&=\mu_B \frac{\mathrm{d}P}{\mathrm{d}\mu_B}-P=A\beta \left(\frac{\mu_B}{\mu_x}\right)^{\beta+1}~.
\end{align}

To generate hybrid EoSs, we employ a two phase approach by matching the pressure and chemical potential of the aforementioned quark model to one of three different hadronic EoSs, namely DD2F~\cite{Typel:2009sy,Alvarez-Castillo:2016oln}, DD2~\cite{Typel:2009sy,Hempel:2009mc} and SFHo~\cite{Hempel:2009mc,Steiner:2012rk}, using a Maxwell construction~\cite{Glendenning:1992vb,Glendenning:2001pe,Hempel:2009vp,Hempel:2013tfa,Constantinou:2023ged}. This scheme leads to a density jump at the density $n_\mathrm{on}$ of size $\Delta n$. A specific choice of $n_\mathrm{on}$, $\Delta n$ and $c_s^2$ together with a given hadronic EoS fixes the parameters $A$, $B$ and $\beta$ by requiring equal pressure and chemical potentials of the hadronic phase at $n_B=n_\mathrm{on}$ and the quark phase at $n_B=n_\mathrm{on}+\Delta n=n_\mathrm{fin}$. 

We test the performance of this parametric approach by reproducing the ensemble of 7 microphysical DD2F-SF EoSs~\cite{Bastian:2020unt} with our scheme (see~\cite{Bauswein:2018bma} for nomenclature). For this, we infer $n_\mathrm{on}$ and $\Delta n$ from the respective EoS. Together with the hadronic DD2F model this directly provides the parameters $A$ and $B$. We obtain $c^2_s$ by fitting Eq.~\eqref{eq:pQ} to each tabulated hybrid EoS in cold, beta-equilibrium composition in the density range from $n_\mathrm{fin}$ to $n_\mathrm{fin}+2\times n_\mathrm{sat}$ using a least squares fit. We provide the inferred parameters in Tab.~\ref{tab:SFfits}. Note that here we have adopted the labeling of the hybrid models from Refs.~\cite{Bauswein:2018bma,Blacker:2020nlq,Blacker:2023afl}.

\begin{figure} 
\centering
\includegraphics[width=1.0\linewidth]{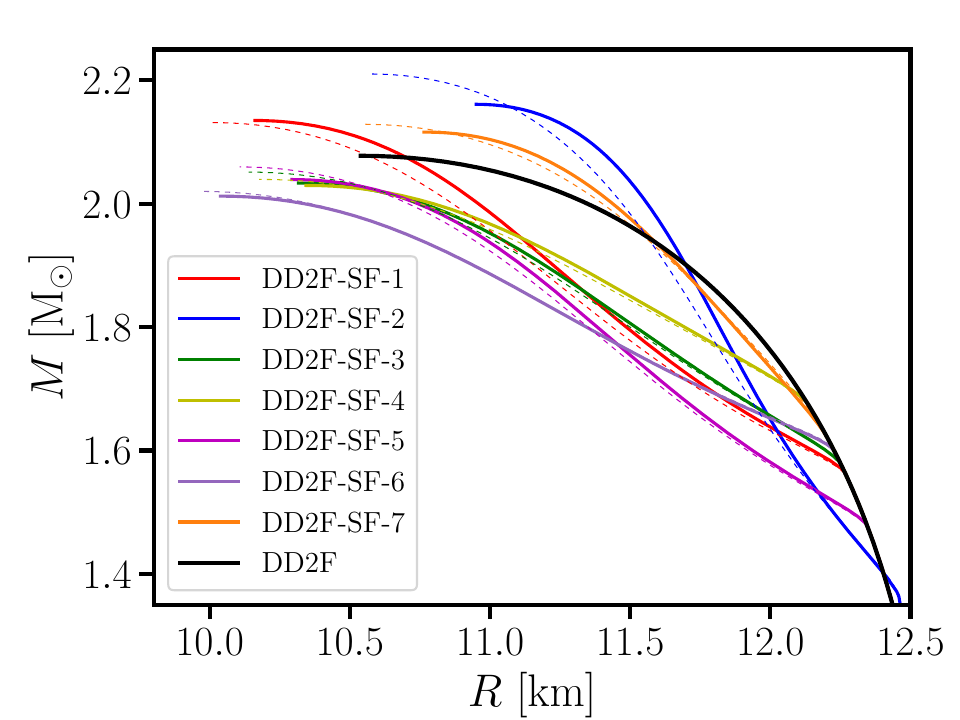}
\caption{Mass-radius curves for all hybrid DD2F-SF EoSs (solid lines). Dashed lines are curves from constant speed of sound fits to the pure deconfined quark phases of these models.}
\label{fig:TOV_DD2F-SF}
\end{figure}

\begin{table}
\caption{Parameters of the constant speed of sound EoSs fitted to the DD2F-SF models }
\begin{tabular}{c | c c c }
\hline\hline
EoSs & $A$ & $B$ & $c_s^2$ \\
 & [MeV/fm$^{3}$] & [MeV/fm$^{3}$] &  \\
\hline\
DD2F-SF-1 & 263.54 & 296.53 & 1.000 \\
DD2F-SF-2 & 194.60 & 206.35 & 0.804 \\
DD2F-SF-3 & 216.71 & 244.75 & 0.793 \\
DD2F-SF-4 & 216.22 & 248.53 & 0.784 \\
DD2F-SF-5 & 215.45 & 239.14 & 0.794 \\
DD2F-SF-6 & 235.94 & 271.54 & 0.835  \\
DD2F-SF-7 & 183.79 & 200.21 & 0.727 \\
\hline
\hline
\end{tabular}
\label{tab:SFfits}
\end{table}

We plot the resulting mass-radius curves of non-rotating NSs for the DD2F-SF EoSs in Fig.~\ref{fig:TOV_DD2F-SF}. Different colors refer to different EoSs. The solid lines mark the original models and dashed lines display results from the fits with the constant speed of sound parametrization. In general, we find that the approach is able to reproduce the mass-radius curves very well with differences in radii below 200~m. Only for the DD2-SF-2 EoS we observe a somewhat larger differences with up to 500~m at high NS masses. This demonstrates that the constant speed of sound parametrization is able to reproduce results from microphysical EoSs with reasonable accuracy.

\section{EffPT scheme and phase boundaries}\label{app:effpt}
In this work, we employ the effPT scheme of Ref.~\cite{Blacker:2023afl} to model the finite temperature regime in merger simulations. It is based on an ideal-gas scheme~\cite{Janka1993,Bauswein:2010dn}, where the pressure and specific energy density are split into a cold and a thermal part as $P=P_\mathrm{cold}+P_\mathrm{th}$ and $\epsilon=\epsilon_\mathrm{cold}+\epsilon_\mathrm{th}$. This scheme is meant to approximate thermal effects if only a barotropic, cold EoS is available with $P_\mathrm{cold}=P_\mathrm{cold}(\rho)$ and $\epsilon_\mathrm{cold}=\epsilon_\mathrm{cold}(\rho)$. The thermal specific energy is directly inferred from solving the hydrodynamic equations as $\epsilon_\mathrm{th}=\epsilon-\epsilon_\mathrm{cold}(\rho)$. The thermal pressure is then calculated with an ideal-gas ansatz 

\begin{align}
    P_\mathrm{th}=(\Gamma_\mathrm{th}-1)\epsilon_\mathrm{th}\rho~,\label{eq:pth}
\end{align}

where the ideal-gas index $\Gamma_\mathrm{th}$ is typically assumed to be constant at all densities and energies.

\begin{figure*}[ht]
\centering
\subfigure[]{\includegraphics[width=0.49\linewidth]{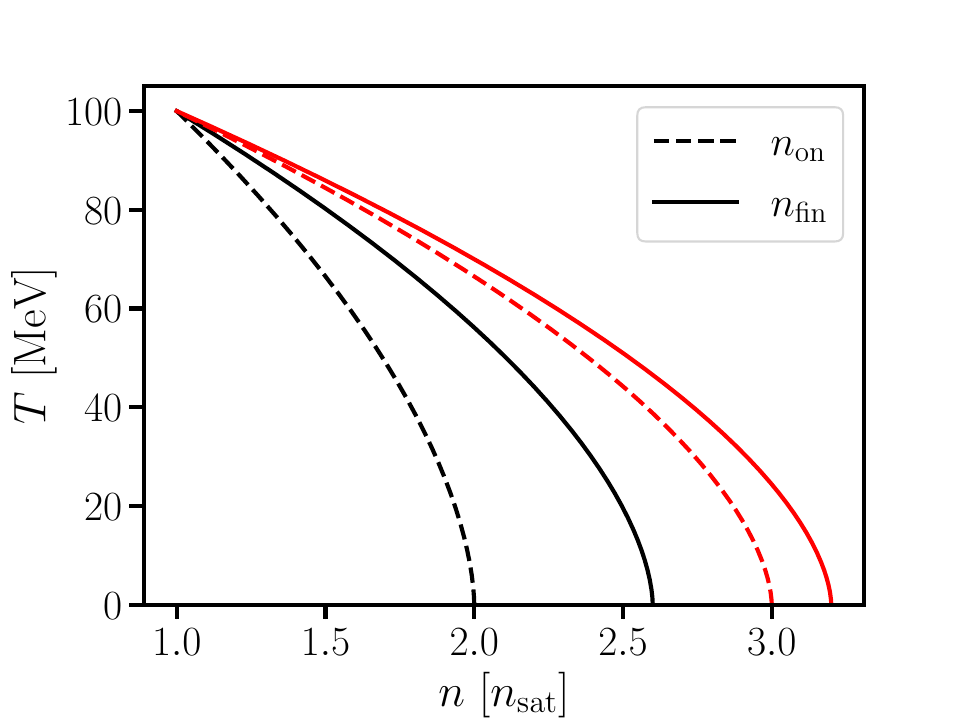}\label{fig:bounds_a}}
\hfill
\subfigure[]{\includegraphics[width=0.49\linewidth]{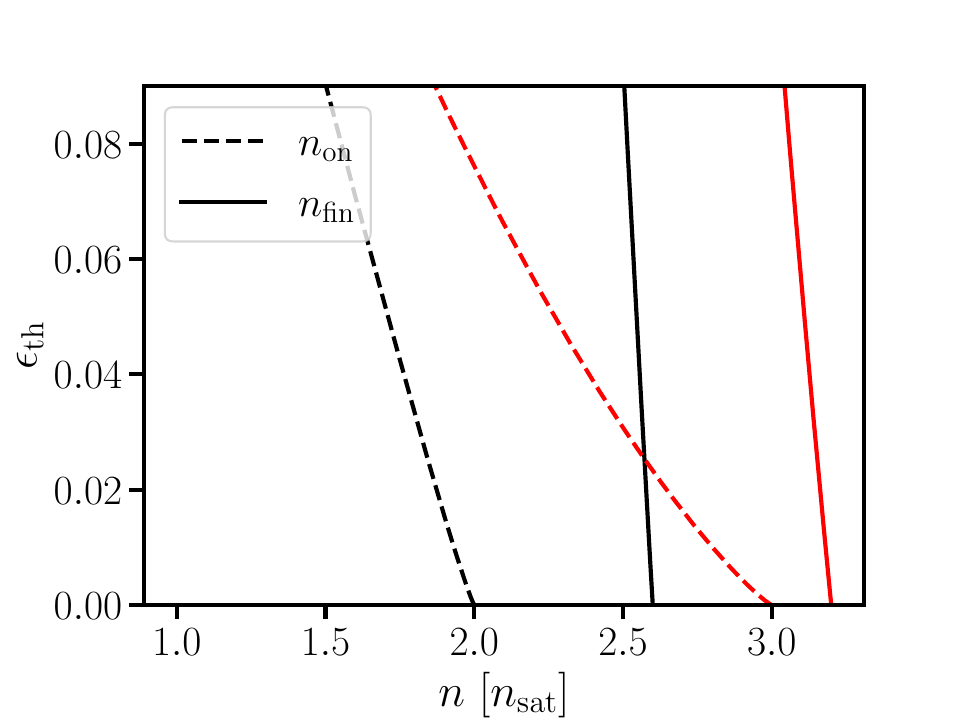}\label{fig:bounds_b}}
\hfill
\caption{(a): Two example phase boundaries for DD2F-based models in the $n$-$T$-plane shown in black and red, respectively. Dashed lines mark the onset {density of quark matter} and solid lines the beginning of the pure deconfined quark matter regime. (b): Same phase boundaries as in (a), but in the $\epsilon_\mathrm{th}$-$n$-plane.}
\label{fig:bounds}
\end{figure*}

The effPT approach extends this scheme by including the changes of the hadron-quark phase boundaries at finite temperatures. In addition to choices of $\Gamma_\mathrm{th}$ for the hadronic and the deconfined quark phase, respectively, the effPT scheme also requires knowledge of the phase boundaries $\rho_\mathrm{on}(\epsilon_\mathrm{th})$ and $\rho_\mathrm{fin}(\epsilon_\mathrm{th})$, which mark the onset and the end of the coexistence phase. As in Ref.~\cite{Blacker:2023afl}, we pick $\Gamma_\mathrm{th}=1.75$ for the hadronic phase and $\Gamma_\mathrm{th}=4/3$ for deconfined quark matter. For both phase boundaries, we assume the following shape

\begin{align}
    \rho_\mathrm{on/fin}(T)=\rho_\mathrm{on/fin,0}-\left(\frac{T}{T_c}\right)^\gamma (\rho_\mathrm{on/fin,0}-\rho_\mathrm{c})~.\label{eq:bounds}
\end{align}
$\rho_\mathrm{on/fin,0}$ refers to the onset/end of the coexistence phase at $T=0$. We fix the parameters by requiring that $\rho_\mathrm{on/fin,0}$ coincide with the onset density and the density jump of the cold hybrid EoS and choosing $\gamma=1.7$. With this choice, the phase boundaries are approximately constant at low temperatures below 20~MeV and shift towards lower densities as the temperature increases (see Fig.~\ref{fig:bounds_a}). From Eq.~\eqref{eq:bounds}, it is clear that both phase boundaries converge towards the value $\rho_c$ at the temperature $T_c$. This is supposed to resemble a behavior of a critical point, where a first-order phase transition changes into a second-order transition and becomes a smooth crossover at higher temperatures. Such a critical point has been hypothesized to exist for the quark-hadron phase transition, see e.g.~\cite{Asakawa:1989bq,Halasz:1998qr,Andronic:2005yp,deForcrand:2008vr}. Motivated by current estimates (see e.g.~\cite{Pandav:2022xxx}), we pick $T_c=100$~MeV and, $\rho_c=\rho_\mathrm{sat}$.

The effPT approach requires the phase boundaries in the $\rho-\epsilon_\mathrm{th}$ plane rather than the $\rho-T$ plane. We obtain these from our assumed phase boundaries in Eq.~\eqref{eq:bounds} by inferring the pressure at the onset phase boundary for all temperatures up to 100~MeV from the respective hadronic EoSs. All three hadronic models we employ in this study are also available as fully temperature-dependent tables. Since we assume Eq.~\eqref{eq:pth} to hold for hadronic matter at all densities, we calculate $\epsilon_\mathrm{th}$ from the pressure assuming $\Gamma_\mathrm{th}=1.75$ along the onset boundary of the coexistence phase. For the beginning of the pure deconfined quark phase, we adopt the same total pressure at a fixed temperature as for the onset density. From this, we determine $\epsilon_\mathrm{th}$ with Eq.~\eqref{eq:pth} by using $\Gamma_\mathrm{th}=4/3$. We remark that our effPT scheme is not able to handle crossovers, since it explicitly assumes a Maxwell construction of the phase transition at all thermal energies $\epsilon_\mathrm{th}$. However, in our simulations we do not encounter the crossover regime since the thermal energies always remain below below the critical point.

We show two example phase boundaries for DD2F-based models in the $n$-$T$- and the $\epsilon$-n-plane in Fig.~\ref{fig:bounds}. As all our phase boundaries end in the same critical point irrespective of $n_\mathrm{on}$ and $\Delta n$ at $T=0$, models with higher $n_\mathrm{on}$ are shifted more strongly towards lower densities at finite temperatures.

Additionally, we slightly modify the effPT scheme for this study compared to Ref.~\cite{Blacker:2023afl}. Due to the 'earlier' {appearance of quark matter} at finite temperature, a density regime exists where cold matter is in the coexistence phase, but hot material can already be composed of pure deconfined quark matter. Correctly determining the pressure of material in this regime requires knowledge of the pure quark EoS at lower densities. This pressure cannot be directly inferred from a cold, tabulated hybrid EoS, as deconfined quark matter is not present at zero temperature in the required density regime. In Ref.~\cite{Blacker:2023afl}, we overcome this problem by using linear extrapolation from the density regime where deconfined quark matter is present at zero temperature to lower densities. Here, we instead infer the pressure of pure quark matter at all densities using Eq.~\eqref{eq:pQ} and then add a thermal component with Eq.~\eqref{eq:pth} and $\Gamma_\mathrm{th}=4/3$.

\section{Additional plots for the baryon chemical potential}\label{app:mub}
\begin{figure*}
\centering
\subfigure[]{\includegraphics[width=0.33\linewidth]{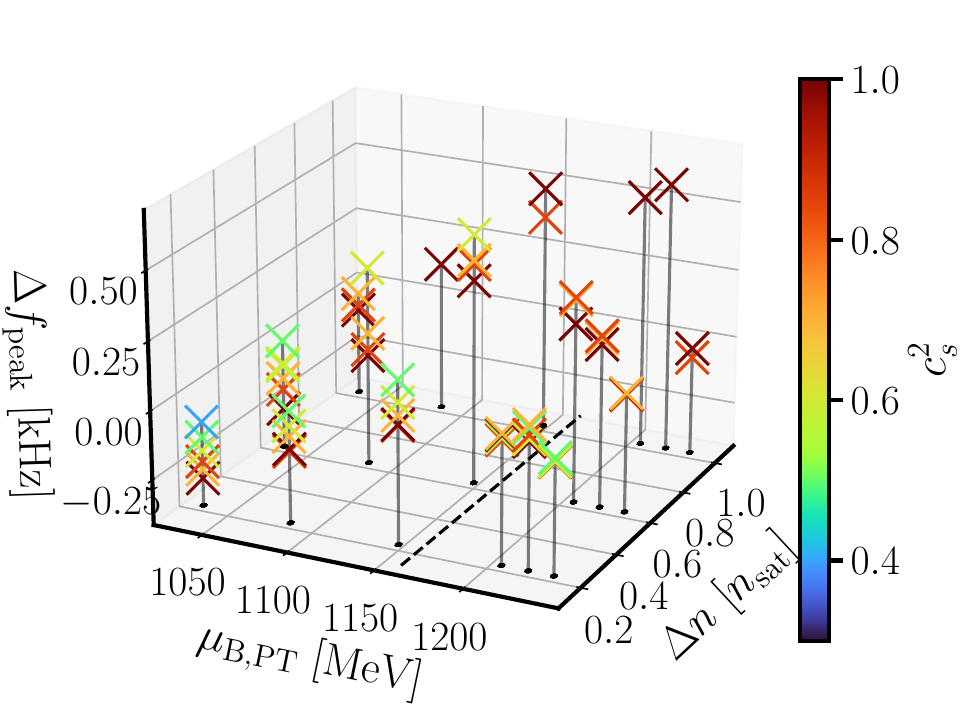}\label{fig:17dresult_a}}
\hfill
\subfigure[]{\includegraphics[width=0.32\linewidth]{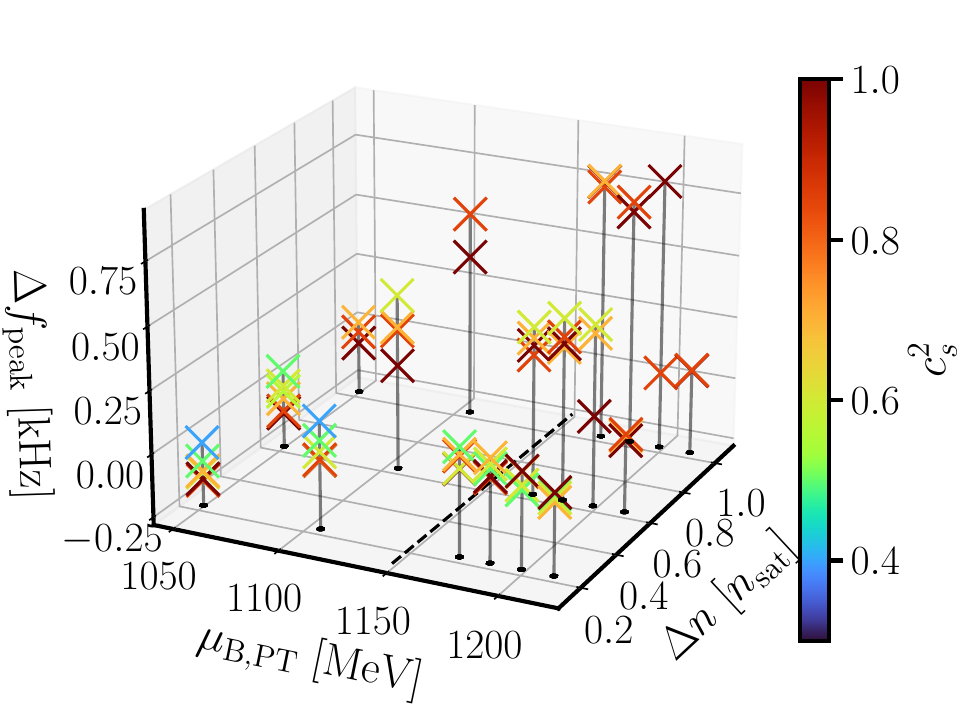}\label{fig:17dresult_b}}
\subfigure[]{\includegraphics[width=0.32\linewidth]{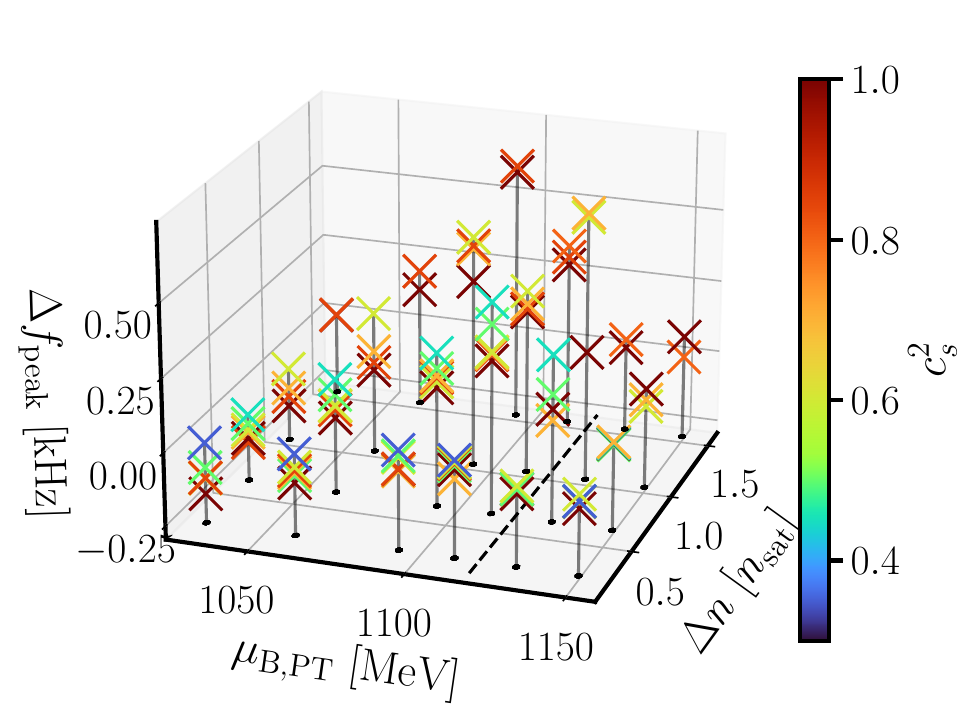}\label{fig:17dresult_c}}
\hfill
\caption{Same as Fig.~\ref{fig:df_DD2F},~\ref{fig:df_DD2} and~\ref{fig:df_SFHO} but with the baryon chemical potential quantifying the onset of the phase transition for models based on SFHO (left), DD2F (middle) and DD2 (right).}
\label{fig:dfmub}
\end{figure*}

\begin{figure*} 
\centering
\subfigure[]{\includegraphics[width=0.33\linewidth]{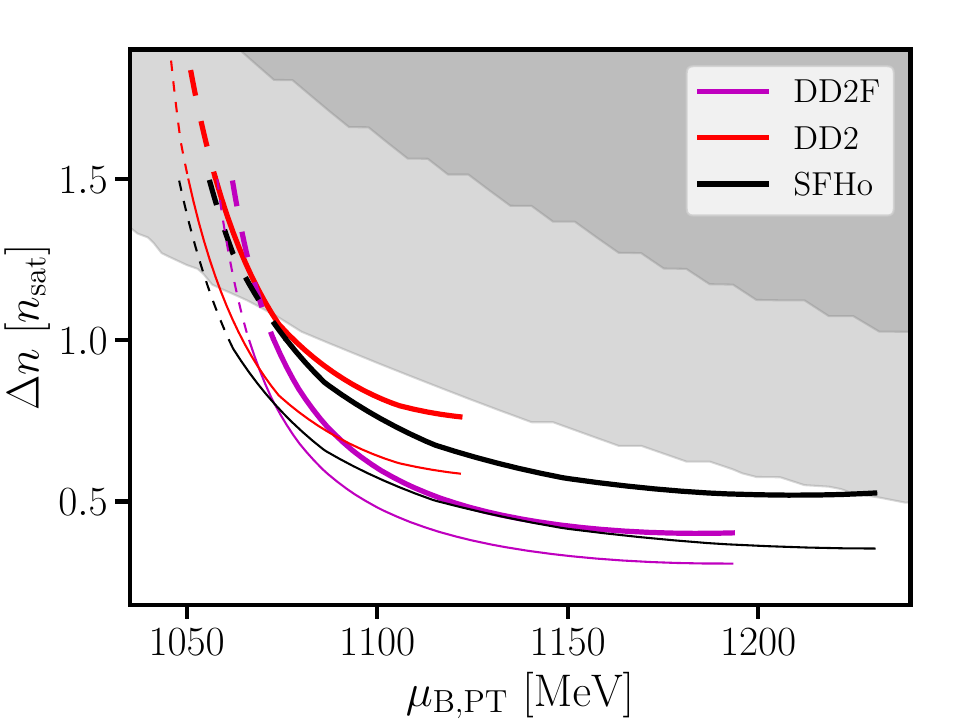}\label{fig:18dresult_a}}
\hfill
\subfigure[]{\includegraphics[width=0.32\linewidth]{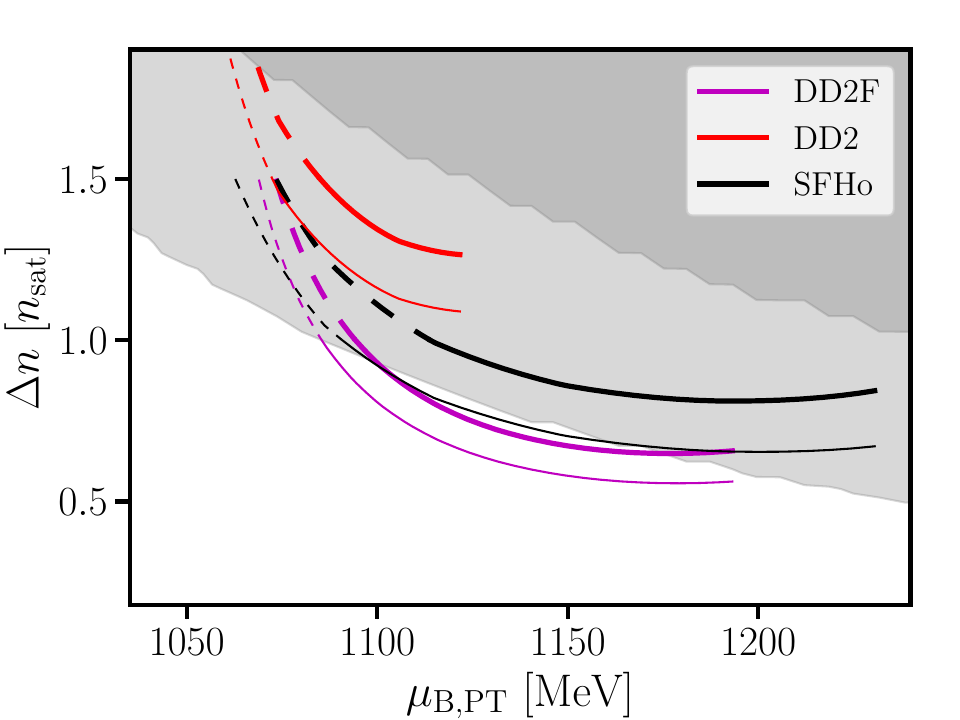}\label{fig:18dresult_b}}
\subfigure[]{\includegraphics[width=0.32\linewidth]{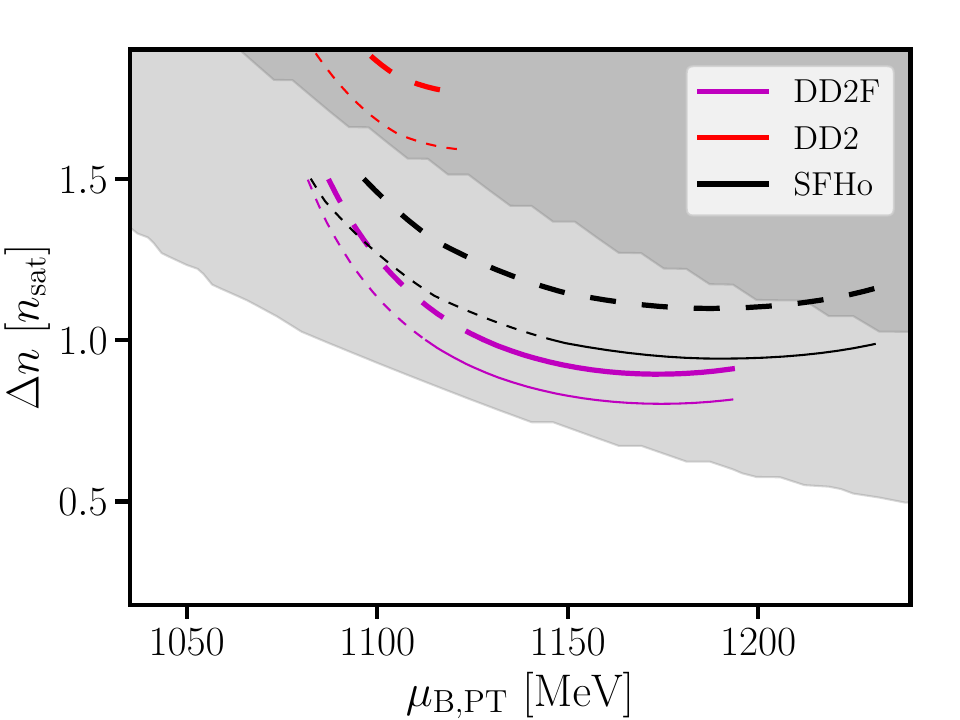}\label{fig:18dresult_c}}
\hfill
\caption{Same as Fig.~\ref{fig:7dresult} but using the baryon chemical potential to illustrate possible constraints on $\mu_\mathrm{B}$ at the phase transition from a determination of $\Delta f_\mathrm{peak}$ assuming $\Delta f_\mathrm{peak}=200$~Hz (left), $\Delta f_\mathrm{peak}=400$~Hz (middle) and $\Delta f_\mathrm{peak}=600$~Hz (right). For simplicity only the gray exclusion regions for DD2F-based models are shown, which slightly deviate from those of hybrid models employing SFHO or DD2.}
\label{fig:18dresult}
\end{figure*}

We report the results for the frequency shift and the constraints on the properties of the phase transition also as function of the baryon chemical potential $\mu_\mathrm{B,PT}$ at the phase transition. Figure~\ref{fig:dfmub} shows the frequency shifts, while constraints on the properties of the phase transition can be read off from Fig.~\ref{fig:18dresult}.

\section{Estimating fit ranges}\label{app:ranges}
As a cross-check we provide an additional estimate of the range of the onset density, where our empirical fit formula (Eq.~\eqref{eq:fitall}) is valid. To this end, we infer the maximum onset density $n_\mathrm{on,max}$ at which a phase transition will still affect $f_\mathrm{peak}$ and beyond which $\Delta f_\mathrm{peak}$ drops to zero. This density can be approximated as the maximum density reached in the merger remnant when employing the underlying hadronic EoS model.

We follow the ideas of Ref.~\cite{Blacker:2020nlq} to calculate this density by exploiting relations between $f_\mathrm{peak}$, $\Lambda$ and the maximum density reached in a merger remnant within the first 5~ms after merger $\rho^\mathrm{max}$. The relation between $f_\mathrm{peak}$ and $\Lambda$ is given in Eq.~\eqref{eq:fpeak_lambda} while $\rho^\mathrm{max}$ can be inferred from $f_\mathrm{peak}$ using the equation~\cite{Blacker:2020nlq}

\begin{align}
\rho^{\mathrm{max}}=a_\rho f_{\mathrm{peak}}^2+b_\rho f_{\mathrm{peak}} +c_\rho \label{eq:rhomaxfpeak}
\end{align}
with $a_\rho=1.689\times 10^{14}$~g/(kHz$^{2}$~cm$^{3}$), $b_\rho=-2.927\times 10^{14}$~g/(kHz~cm$^{3}$), $c_\rho=2.837\times 10^{14}$~g/(~cm$^{3}$). By using Eqs.~\eqref{eq:fpeak_lambda} and \eqref{eq:rhomaxfpeak}, we can estimate $\rho^\mathrm{max}$ for a hadronic EoS based on the tidal deformability of the inspiraling stars. 

Additionally, we need to take into account that small amounts of deconfined quark matter do not affect $f_\mathrm{peak}$ much. To account for this, Ref.~\cite{Blacker:2020nlq} introduced an additional parameter $\Delta M$. If for a given system mass $M_\mathrm{tot}$ deconfined quark matter is present in a system but $f_\mathrm{peak}$ is not yet affected, then a fiducial system $X$ with a larger mass $M_X=M_\mathrm{tot}+\Delta M$ will show clear signatures of quark matter. The tidal deformability $\Lambda^{X}$ of this fiducial system can be estimated as $\Lambda^{X}=\Lambda-5.709\frac{\Lambda}{M_{\mathrm{tot}}}\times \Delta M$~M$_{\odot}$. 

Using $\Lambda_{1.35}$ of our three underlying hadronic EoSs and picking $\Delta M=0.2~M_\odot$ (compare Ref.~\cite{Blacker:2020nlq}), we obtain $n_\mathrm{on,max}=2.86\times n_\mathrm{sat}$, $n_\mathrm{on,max}=2.05\times n_\mathrm{sat}$ and $n_\mathrm{on,max}=3.43\times n_\mathrm{sat}$ for DD2F, DD2 and SFHo, respectively. These values are in good agreement with the ranges found in Sec.~\ref{sec:results}.

\end{appendix}

\end{document}